\documentclass[12pt,preprint]{aastex}

\usepackage{enumerate}
\usepackage{float}
\usepackage{placeins}
\usepackage{color}
\usepackage{epsfig}
\usepackage{graphicx}
\usepackage{amsmath}
\usepackage[utf8x]{inputenx}
\usepackage{natbib}
\usepackage{footnote}
\usepackage{pdflscape}
\usepackage{aas_macros}
 \usepackage{hyperref}

\def\HII{{\ion{H}{2}}}

\def\4959_5007{[\ion{O}{3}]~$\lambda \lambda$4959,5007}
\def\OIII49595007{[\ion{O}{3}]~$\lambda \lambda 4959,5007$}
\def\ratioR23{([\ion{O}{2}]~$\lambda \lambda$3727,9 + [\ion{O}{3}]~$\lambda\lambda$4959,5007)/H$\beta$}
\def\R23{${\rm R}_{23}$}
\def\dS23{${\rm S}_{23}$}

\def\lOII{[\ion{O}{2}]~$\lambda\lambda$3737,9}
\def\lOIII{[\ion{O}{3}]~$\lambda$5007}
\def\lNII{[{\ion{N}{2}}]~$\lambda$6584}
\def\SIIl{[\ion{S}{2}]~$\lambda \lambda$6717,31}

\def\NII{[{\ion{N}{2}}]}

\def\ratioS23{([\ion{S}{2}]~$\lambda \lambda$6717,31 +[\ion{S}{3}]~$\lambda\lambda$9069,9532)/H$\beta$}

\def\SII{[{\ion{S}{2}}]}

\def\O4363{[{\ion{O}{3}}]~$\lambda$4363}
\def\OIII{[{\ion{O}{3}}]}

\newcommand{\lzifu} {{\scshape lzifu}}
\newcommand{\ppxf} {{\scshape ppxf}}

\shorttitle{The S7 Survey}
\shortauthors{Dopita et al.}

\begin{document}

\title{Probing the Physics of Narrow Line Regions in Active Galaxies II: \\ The Siding Spring Southern Seyfert Spectroscopic Snapshot Survey (S7)}

\author{
Michael A. Dopita\altaffilmark{1,2}, Prajval Shastri\altaffilmark{3}, Rebecca Davies\altaffilmark{1}, Lisa Kewley\altaffilmark{1,4},  Elise Hampton\altaffilmark{1}, Julia Scharw\"achter\altaffilmark{5}, Ralph Sutherland\altaffilmark{1}, Preeti Kharb\altaffilmark{3}, Jessy Jose\altaffilmark{3}, Harish Bhatt\altaffilmark{3}, \\ S. Ramya  \altaffilmark{3}, Chichuan Jin\altaffilmark{6}, Julie Banfield\altaffilmark{7}, Ingyin Zaw\altaffilmark{8}, St\'ephanie Juneau\altaffilmark{9}, \\ Bethan James\altaffilmark{10}  \& Shweta Srivastava\altaffilmark{11}
}
\email{Michael.Dopita@anu.edu.au}

\altaffiltext{1}{RSAA, Australian National University, Cotter Road, Weston Creek, ACT 2611, Australia}
\altaffiltext{2}{Astronomy Department, King Abdulaziz University, P.O. Box 80203, Jeddah, Saudi Arabia}
\altaffiltext{3}{Indian Institute of Astrophysics, Koramangala 2B Block, Bangalore 560034, India }
\altaffiltext{4}{Institute for Astronomy, University of Hawaii, 2680 Woodlawn Drive, Honolulu, HI, USA}
\altaffiltext{5}{LERMA, Observatoire de Paris, CNRS, UMR 8112, 61 Avenue de l'Observatoire, 75014, Paris, France}
\altaffiltext{6}{Qian Xuesen Laboratory for Space Technology, Beijing, China }
\altaffiltext{7}{CSIRO Astronomy \& Space Science, P.O. Box 76, Epping NSW, 1710 Australia }
\altaffiltext{8}{New York University (Abu Dhabi) , 70 Washington Sq. S, New York, NY 10012, USA }
\altaffiltext{9}{CEA-Saclay, DSM/IRFU/SAp, 91191 Gif-sur-Yvette, France}
\altaffiltext{10}{Institute of Astronomy, Cambridge University, Madingley Road, Cambridge CB3 0HA, UK }
\altaffiltext{11}{Astronomy and Astrophysics Division, Physical Research Laboratory, Ahmedabad 380009, India}

\begin{abstract}
Here we describe the \emph{Siding Spring Southern Seyfert Spectroscopic Snapshot Survey} (S7) and present results on 64 galaxies drawn from the first data release. The S7 uses the Wide Field Spectrograph (WiFeS) mounted on the ANU 2.3m telescope located at the Siding Spring Observatory to deliver an integral field of $38\times25$~ arcsec at a spectral resolution of $R=7000$ in the red ($530-710$nm), and $R=3000$ in the blue ($340-560$nm). {From these data cubes we have extracted the Narrow Line Region (NLR) spectra from a 4 arc sec aperture centred on the nucleus.}   We also determine the H$\beta$ and \lOIII\ fluxes in the narrow lines,  the nuclear reddening, the reddening-corrected relative intensities of the observed emission lines, and the H$\beta$ and \lOIII\ luminosities {determined from spectra for which the stellar continuum has been removed.} We present {\bf a set of } images of the galaxies in \lOIII,  \lNII\ and H$\alpha$ which serve to delineate the spatial extent of the extended narrow line region (ENLR) and {\bf also to} reveal the structure and morphology of the surrounding \HII\ regions. Finally, we provide a preliminary discussion of  those Seyfert~1 and Seyfert~2 galaxies which display coronal emission lines in order to explore the origin of these lines. \end{abstract}


\section{Introduction}\label{sec:intro}
It has been understood for many years that massive black holes are ubiquitous in the centres of the more massive disk and elliptical galaxies, and that there exists an intimate connection between the black hole mass and the host galaxy measured either through the bulge mass, luminosity or velocity dispersion  \citep{Magorrian98, Ferrarese00, Gebhardt00, Tremaine02, McConnell13}. Furthermore, it has been established these relationships were already in place by $z\sim 2$ \citep{Bennert11}, demonstrating that the build up of mass in the host galaxy bulges was contemporaneous with the growth of their central black holes.

Our understanding of the more recent mass feeding of these nuclear black holes remains somewhat sketchy. We can measure the luminosity of the AGN in the local universe, but this provides the product of the black hole mass and the Eddington fraction. In terms of the electron scattering opacity the Eddington fraction is defined as $f_{\rm Edd.} = L_{\rm BH}/\left(1.25\times10^{38} {\rm erg~s^{-1}}\left[M_{\rm BH}/M_{\odot}\right]\right)$. In practice, it is very difficult to separate these two variables of black hole mass and Eddington fraction. In addition, the issues of orientation and obscuration sometimes render it difficult to even estimate the luminosity of the central engine, unless a very long baseline spectral energy distribution (SED) is available. According to the ``standard'' unified model of AGN \citep{Antonucci90,Antonucci93} and its extensions -- which attempt to account for the effect of the Eddington fraction of the accretion rate \citep{Dopita97} -- the Seyfert~1 galaxies are seen pole-on relative to the accretion disk, and these display very broad permitted lines originating in rapidly moving gas close to the central engine. In the Seyfert~2 galaxies, the thick accretion disk obscures the central engine, and an Extended Narrow Line Region (ENLR) often confined within an ``ionisation cone" is observed. In this geometry, the ENLR can be readily observed even when the central engine is very heavily obscured at optical wavelengths \citep{Kewley01}. The fundamental problem with the original unified model with a thick accretion disk is that, if continuous, such a disk could no support itself against collapse in the vertical direction. The more generally accepted model is now the clumpy torus model \citep{Nenkova02,Elitzur06,RamosAlmeida09}, but this more physically motivated model still operates to obscure the central engine over a range of angles and also to confine the escape of the EUV radiation to the polar directions.

The properties of the ENLR can provide vital clues about the nature of the central black hole, and the mechanisms which produce the extreme UV (EUV) continuum. Seyfert galaxies are known to occupy a very restricted range of line ratios when plotted on the well-known BPT diagram \citep{Baldwin81} which plots [N II] $\lambda$6584/H$\alpha$ vs. [O III] $\lambda$5007/H$\beta$ or alternatively, using the other diagrams introduced by \citet{Veilleux87} involving either the  [S II] $\lambda$6717,31/H$\alpha$ ratio or the [O I] $\lambda$6300/H$\alpha$ ratio in the place of the [N II] $\lambda$6584/H$\alpha$ ratio. It now seems clear that this is because the ENLR is, in general, radiation pressure dominated  \citep{Dopita02,Groves04a, Groves04b}. In this model, radiation pressure (acting upon both the gas and the dust) compresses the gas close to the ionisation front so that at high enough radiation pressure, the density close to the ionisation front scales as the radiation pressure, and the local ionisation parameter ($U$, the ratio of the density of ionizing photons to the ion density) in the optically-emitting ENLR becomes constant. This results in an optical ENLR spectrum which is virtually independent of the input ionisation parameter. For dusty ENLRs the radiation pressure comes to dominate the gas pressure for $\log U \gtrsim -2.5$, and the emission spectrum in the optical becomes invariant with the input ionisation parameter for $\log U \gtrsim -0.5$. In this condition, the observed density in the ENLR should drop off in radius in lockstep with the local intensity of the radiation field; $n_e \propto r^{-2}$, and the EUV luminosity can be inferred directly from a knowledge of the density and the radial distance.

In the first paper of this series, \citet{Dopita14} applied these ideas to a test-case example Seyfert, NGC~5427, in order to determine how well the EUV spectrum, luminosity, and black hole mass can be determined from an analysis of the narrow line spectrum of the nucleus and its associated ENLR. In this paper they utilised an idea originally put forward by \citet{Evans87}, namely, to use an analysis of the \HII\ regions surrounding the AGN to constrain the chemical abundance of the ENLR. By thus eliminating chemical abundance as a free variable, the gross features of the EUV spectrum (between 13.6 and $\sim 150$eV ) can be inferred by a method similar to the energy balance or Stoy technique which has long been used to estimate the effective temperature of stars in planetary nebulae \citep{Stoy33,Kaler76,Preite-Martinez83}. This relies on the fact that as the radiation field becomes harder, the heating per photoionisation increases, and the sum of the fluxes of the forbidden lines becomes greater relative to the recombination lines. Furthermore, additional constraints are available because individual line ratios are sensitive in different ways to the form of the EUV spectrum. These properties can be exploited to infer the form of the EUV spectrum, and application of this technique  will be the subject of future papers in this series.

Apart from the emission line spectrum, the dynamical structure of the ENLR is also of great interest. Most ENLR show velocity dispersions of up to a few hundred km~s$^{-1}$. It is not yet clear what fraction of this velocity dispersion is due to outflows powered originally by circum-nuclear starbursts, what is the contribution to the velocity dispersion of radiatively-driven outflows \citep{Cecil02,Dopita02,Mullaney09}, or what fraction is generated by the cocoon shocks powered by the overpressure of relativistic plasma derived from the radio jets \citep{Bicknell98,Tadhunter14}.

At sufficiently great a distance, the spectra of the ENLR and the \HII\ regions become mixed together within a single resolution element. This mixing has been investigated by \citet{Scharwachter11} and \citet{Dopita14}, and further quantified by \citet{Davies14b,Davies14a}. From the BPT diagram \citep{Baldwin81} and the \citet{Veilleux87} diagnostics, it is possible to both clearly define the zone of influence of the AGN and to quantify the total luminosity of the ENLR. A future paper will investigate this mixing in the case of the Seyferts with ENLR presented here.

The other major class of AGN in the local universe are the low ionisation nuclear emission line regions (LINERs), first defined as a class by  \citet{Heckman80}. The host galaxies of these objects are generally large ellipticals, although they are also found in some nearby spirals such as M~81. We now understand that LINERs represent AGN with low Eddington fractions \citep{Kewley06}, but we are still a long way from understanding why this should produce their characteristic LINER spectrum in the optical. The most likely mechanisms are shocks \citep{Koski76, Fosbury78, Baldwin81} or else photoionisation by either a power-law or a thermal Bremstrahhlung continuum with a low ionization parameter \citep{Ferland83}.

Low level LINER emission has now been detected in a large fraction of elliptical galaxies \citep{Phillips86, Veron-Cetty86, Ho96,Ho97}. Surveys by \citet{Oconnell78} and \citet{Heckman80} confirmed a strong correlation between bright LINER emission and powerful compact nuclear radio sources. In most of these bright LINERs, virtually all of the radio emission comes from flat-spectrum compact self-absorbed synchrotron sources \citep{Condon78}.

Recently it has become clear that  the frequently-detected ``extended'' LINER emission is not necessarily of the same origin as the true ``nuclear'' LINER activity. \citet{Maoz98} finds that the LINER class is not homogenous in its UV properties -- some objects exhibiting strong emission lines, while others display a UV spectrum that is consistent with an old stellar population. Indeed, for ``extended'' LINER emission \citet{Yan12} proved that the extended emission is inconsistent with ionisation from a central object, and that most likely post-AGB stars are required to provide the ionisation. This idea was originally proposed by \citet{Binette94}, and has been recently developed in the light of CALIFA  integral field spectroscopic data by \citet{Singh13}. Such Low Ionisation Extended Regions perhaps should be more appropriately referred to as ``LIERs"!

From the above discussion it should be clear that a great deal of interesting physics can be derived from a survey of ENLR of nearby Seyferts and of LINER galaxies, provided that this survey has adequate spectral resolution to investigate the dynamics, adequate spectral coverage to provide complete strong line diagnostics and to allow unambiguous photoionisation modelling, and adequate spatial resolution to both resolve the spatial and dynamical structure of the ENLR and to isolate individual \HII\ region complexes. This clearly defines a survey which can only be undertaken using an Integral Field Unit (IFU). 

Up to the present, Seyfert and LINER galaxies have been identified mainly through single-aperture spectroscopic surveys, of which the most notable is the SDSS survey. In its seventh data release DR7 \citep{Abazajian09} the SDSS spectroscopic catalogue covers over $10^4$ deg$^2$ of the high-latitude sky and contains many thousands of Seyferts and LINERS as well as $\sim 10^5$ quasars. The physical properties of the Seyferts and LINERs in this catalogue have recently been effectively examined by \citet{Zhang13}.

Surveys probing the extent of the ENLR of Seyfert galaxies are rather sparse. Most of these are derived from images taken in the \lOIII\ emission line --which is very strong in the ENLR -- and often using H$\alpha$ + \lNII\ images as well. This allows one to distinguish the \HII\ regions from the ENLR \citep{Pogge89, Haniff88, Wilson88, Evans96,Davies14a,Davies14b}. The most extensive ground-based surveys of both the extent and morphology of Seyferts are still those of \citet{Mulchaey96a,Mulchaey96b}. \citet{Falcke98} used the WFPC2 imager on the Hubble Space Telescope to identify extended \lOIII\ emission in seven Seyfert 2 galaxies, as did \citet{Schmitt03a,Schmitt03b} in a sample of 60 Seyfert galaxies (22 Sy1 and 38 Sy 2 galaxies), selected based on their far-infrared properties. Finally, we should mention the important UV survey of \citet{MunozMarin07}, which probed the central regions of 75 Seyfert galaxies imaged in the near-UV with the Advanced Camera for Surveys of the Hubble Space Telescope at an average resolution of $\sim10$ pc.

Here, we present initial results from the first data release of the \emph{Siding Spring Southern Seyfert Spectroscopic Snapshot Survey} -- S7. This comprises an integral field survey of over 130 galaxies in total, of which 64 galaxies are included in this first data release. In this paper we concentrate on a presentation of the narrow-band images of the Seyfert galaxies in our sample, and we identify the objects showing pronounced ENLR and/or circum-nuclear star formation activity. We also present the nuclear spectra of all objects extracted from a 4 arc~sec diameter aperture, and present the measured reddening-corrected emission line fluxes between \lOII\ and \SIIl\ .  In Section \ref{S7} we describe the characteristics of this survey, in Section \ref{Obs.} the observational data set and  in Section \ref{reduction} the reduction techniques used. Our results are given in Section \ref{results}.  In section \ref{nuc_spectra} we present the results of fitting stellar continua and a 3-component emission line model to our spectra. This allows the extraction of the nuclear properties. In Section \ref{coronal} we examine the systematics of both Seyfert 1 and Seyfert 2 galaxies with coronal line emission.  The results from the narrow-band images extracted from the data cubes are presented in Section \ref{Imageresults}, in which we note the angular and physical sizes of the ENLR, and probe the relationship between the ENLR and the \HII\ regions. The results on individual galaxies are given in Section \ref{Notes} Finally in Section \ref{conclusions} we present our conclusions. In this paper, we assume $H_0 = 71$~km~s$^{-1}$Mpc$^{-1}$, following the 7 year WMAP results (Larson et al. 2011).

\section{ An Integral Field AGN Survey}\label{S7}
\subsection{The S7 Survey}
The S7 survey is an integral field survey in the optical of $\sim 140$ southern Seyfert and LINER galaxies. It uses the Wide Field Spectrograph (WiFeS) mounted on the Nasmyth focus of the ANU 2.3m telescope. This instrument is described in  \citet{Dopita07}, and its performance is discussed in \citet{Dopita10}. WiFeS provides data cubes over a field of $38\times25$ arcsec at a spatial resolution of 1.0 arcsec It covers the waveband $340-710$~nm with the unusually high resolution of $R=7000$ in the red ($530-710$~nm), and $R=3000$ in the blue ($340-570$~nm). The typical throughput of the instrument (top of the atmosphere to back of the detector) is $20-35$\% \citep{Dopita10}, which provides an excellent sensitivity to faint low surface brightness ENLR features, while the high resolution ($\sim 50$km~s$^{-1}$) in the red enables the different velocity components of emission lines to be clearly separated.

The S7 survey is by no means a complete survey of southern Seyfert, LINERs and other active galaxies but rather, offers a reasonably representative sample of radio-detected AGN in the nearby universe. The objects for the S7 were selected from the \citet{VC06} catalogue of active galaxies, which remains the most comprehensive compilation of active galaxies in the literature. Since, in the future,  we wish to investigate the interaction of the bipolar plasma jets with the Narrow Line Region and the ISM of the host galaxy, we limited the sample to galaxies with radio flux densities high enough to permit radio aperture synthesis observations.  We adopted the following selection criteria:
\begin{itemize}
\item Declination  $<10\degr$ to avoid observations being made at too great a zenith distance. At a zenith distance of 60 degrees, the atmospheric dispersion between 3500 and 7000\AA\ is nearly 6~ arcsec,which would seriously compromise the effective field coverage.
\item Galactic latitude $>|20|$ degrees (with a few exceptions in the case of objects known to have an associated ENLR ). This galactic latitude requirement is set  to avoid excessive galactic extinction.
\item Radio flux density at 20cm $\gtrsim20$mJy for those targets with declination N of -40 deg, which have NVSS measurements, and
\item Redshift $<0.02$. This criterion ($D< 80$ Mpc) ensures that the spatial resolution of the data is better than 400 pc arc~sec$^{-1}$, sufficient to resolve the ENLR, and to ensure that the important  diagnostic [S~II] lines are still within the spectral range of the WiFeS high-resolution red grating.
\end{itemize}

\subsection{The Observations}\label{Obs.}
Each target was observed with the nucleus centred in the WiFeS aperture, at a position angle either close to the major axis of the galaxy, or along the axis of the radio jet and/or the ENLR, where this axis was known. {The observing strategy was as follows. Galaxies were observed in pairs, which were chosen to be separated on the sky by no more than about $15\degr$. In each galaxy we selected a nearby blank sky reference region which could then be used for the purpose of sky subtraction in either galaxy}. 

A typical complete series of exposures comprising a single observing sequence would be $1\times$ Sky region\#1, $3\times$ Galaxy\#1, $1\times$ Sky region\#1,  Bias, Copper-Argon Arc Calibration, $1\times$ Sky region\#2, $3\times$ Galaxy\#2, and $1\times$ Sky region\#2. The exposure times of the individual WiFeS frames ranged from 800s to 1000s, depending on the observing conditions and the length of the night. Observing in this manner allowed us to combine the first two sky frames with the third sky frame to provide the sky reference for Galaxy \#1, while the second sky frame was combined with the third and the fourth to provide the sky reference for Galaxy \#2, giving an on-target observation efficiency of close to 60\%.

The absolute photometric calibration was made using the STIS spectrophotometric standard stars \footnote{Available at : \newline {\tt www.mso.anu.edu.au/~bessell/FTP/Bohlin2013/GO12813.html}}. In addition a number of B-type telluric standards were observed to correct for the OH and H$_2$O telluric absorption features in the red. The separation of these features by molecular species allowed for a more accurate telluric correction which accounted for night to night variations in the column density of these two species. In addition a set of calibration flat fields, and twilight sky flats were taken in each of the five observing runs in which the data was collected.

\begin{figure}[htb!]
\begin{centering}
\includegraphics[scale=0.85]{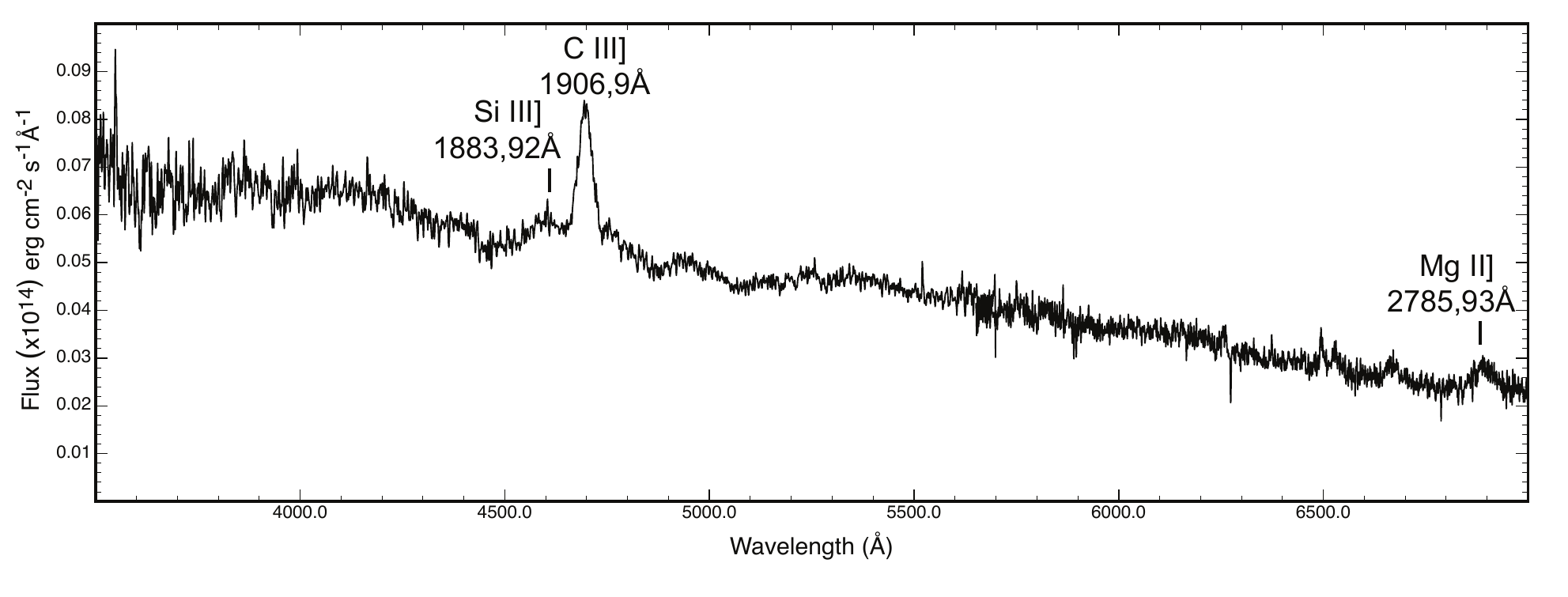}
\end{centering}
\caption{The measured spectrum and the line identifications for PKS 0056-572. We conclude that this object is a QSO at a redshift of $z=1.46$.}\label{fig:fig1}
\end{figure}

In Table \ref{table:Obs} we present the observing log {including the date of observation, the total exposure time in all three sub-exposures and the mean seeing during each observation as measured in the auto guider. This Table also} includes the classification of the nuclear activity, and comparison with the type given in \citet{VC06}. A number of mis-matches are evident. Most spectacular is the case of PKS 0056-572, which was classified as a Seyfert 1, but is in fact a QSO at a redshift of $z=1.46$ (see Figure \ref{fig:fig1}). In addition a number of LINER galaxies are classified in the \citet{VC06} catalog as S2. One galaxy, 3C278 has no detectable emission features, and appears to be simply an Elliptical galaxy on the basis of its stellar continuum. In Table \ref{table:Obs} we have also noted the cases where coronal emission features are evident. The [\ion{Fe}{7}] 6087, 5721 \AA\ lines are visible in all of these, but in the objects with strong high-excitation coronal features, the [\ion{Fe}{14}] 5303\AA\ and  [\ion{Fe}{10}] 6374\AA\ lines are also prominent. Table \ref{table:Obs}  also provides information on the observed extent of the ENLR, and notes those cases amongst the Seyfert 2 galaxies in which the fitting procedure (see Section \ref{fits}, below) indicated the presence of an underlying broad component to the H$\alpha$ profile, which might be the signature of a highly extinguished or hidden Broad Line Region (BLR).

\section{Data reduction}\label{reduction}

The data were reduced using the standard {\tt PyWiFeS} pipeline written for the instrument and fully described in \citet{Childress14}. In brief, this produces a summed data cube for each object wavelength calibrated and evenly sampled in wavelength, sensitivity corrected in both the spatial and spectral directions (including telluric corrections), photometrically calibrated using standard stars, and from which cosmic ray events have been removed. The red and a blue arm data were reduced independently, producing spectral data cubes regularly sampled at 0.77\AA\ in the blue, and 0.44\AA\ in the red.

In this paper we will restrict ourselves to an extraction of emission line images from the whole data cube, and an examination of the nuclear spectra. However, future papers will examine the issues of mixing between the ENLR and \HII\ regions within a single spaxel, the ENLR structure and dynamics, the chemical abundances derived from the \HII\ regions, and the constraints that can be placed on the nature of the central engine from the spectroscopy.

\subsection{Images}\label{Images}
We have extracted continuum-subtracted emission line images from the data cubes using {\tt QFitsView v3.1 rev.741}. \footnote{{\tt QFitsView v3.1} is a FITS file viewer using the QT widget library and was developed at the Max Planck Institute for Extraterrestrial Physics by Thomas Ott.}  Our objective was to separate the emission line objects in the field by their excitation mechanism. We therefore combined  the \lOIII\ image (blue channel) with the \lNII\ image (green channel) and placed the H$\alpha$ image in the red channel. With this scheme, the high metallicity \HII\ regions found in the vicinity of Seyfert nuclei appear red, gold, or sometimes yellow since their \lOIII\ emission is generally weak, and the \lNII\ line is usually weaker than H$\alpha$. The Seyfert nuclei, on the other hand, appear blue, turquoise or sometimes green since their \lOIII\ emission is very strong, and in many cases the \lNII\ emission is as strong or even stronger than H$\alpha$.

\subsection{Circum-nuclear Spectra}\label{fits}
{From the data cube, we have extracted the circum-nuclear spectra of each galaxy from a fixed 4 arcsec diameter aperture using {\tt QFitsView}. This aperture - typically of order 1kpc at the galaxy, was chosen so as to provide the integrated nuclear fluxes of all galaxies, regardless of seeing conditions. } Any remaining residual of the night sky lines of [\ion{O}{1}] $\lambda$ 5577.3, [\ion{O}{1}] $\lambda$ 6300.3 and  [\ion{O}{1}]$ \lambda$ 6363.8 were removed by hand. In addition - in the case of the small fraction of data obtained in non-photometric conditions - a scaling factor was sometimes applied to the red spectrum using the relative scaling determined in the spectral overlap region between 5500 and 5600\AA. The data were smoothed by a boxcar function with a 1:3:5:3:1 weighting function to remove noise within a given resolution element, and then shifted to rest wavelength using the measured wavelength of the \lNII\ line.

\section{Results} \label{results}
\subsection{Circum-nuclear Spectra}\label{nuc_spectra}
We used the IFS toolkit \lzifu\ (Ho et al. 2014 in prep.) to derive gas and stellar kinematics from the nuclear spectra. \lzifu\ uses the penalized pixel-fitting routine \citep[\ppxf\;][]{Cappellari04} to perform simple stellar population (SSP) synthesis fitting to model the continuum, and fits the emission lines as Gaussians using the Levenberg-Marquardt least-squares technique \citep{Markwardt09}. We employ the theoretical SSP libraries from \citet{Gonzalez05} assuming the Padova isochrones. For the gas velocity and velocity dispersion, we simultaneously fit up to three Gaussians to  each of the optical emission lines and constrain each of the components to have the same velocity and velocity dispersion for every emission line. The exception is in the case of Seyfert 1 galaxies, where the broad component is fit to only the Balmer and Helium recombination lines. We fix  the ratios [\ion{O}{3}]$ \lambda \lambda 4959/5007$ and [\ion{N}{2}]$ \lambda \lambda 6548/6584$ to their theoretical values given by quantum mechanics \citep{ADU}. 

The choice of the number of Gaussians to fit is driven by the quality of the fit. First one, then two and if needed, three Gaussian profiles are fit to the narrow lines. If there is no improvement in the residuals by the addition of another component, the number of Gaussians to be used is fixed at the lower number.  In the case of the Seyfert 1 galaxies, either one broad component, or a broad plus intermediate width component are required to fit the broad line profile, which restricts the quality of the fit that can be obtained in the narrow line region. In general, the coronal emission in those objects which display these lines are well fit by the intermediate component.

The reddening is determined independently for the stellar continuum component and the emission lines. Since the measured emission line fluxes determined for the higher members of the Balmer series may be in error due to the competition between the emission in the line and the absorption in the underlying stellar continuum (which is very sensitive to the age of the younger stellar components), we have avoided the use of the higher members of the Balmer series in the extinction determination, and have used only the H$\alpha$/H$\beta$ ratio to determine A$_{\rm V}$ using the formulation given in \citet{Vogt13}. We assumed the  H$\alpha$/H$\beta$ ratio to be its theoretical value at a temperature of 10000K and a density of $10^3$cm$^{-3}$ -- which is 2.86 \citep{ADU}.

\begin{figure*}[htb!]
\begin{centering}
\includegraphics[scale=0.85]{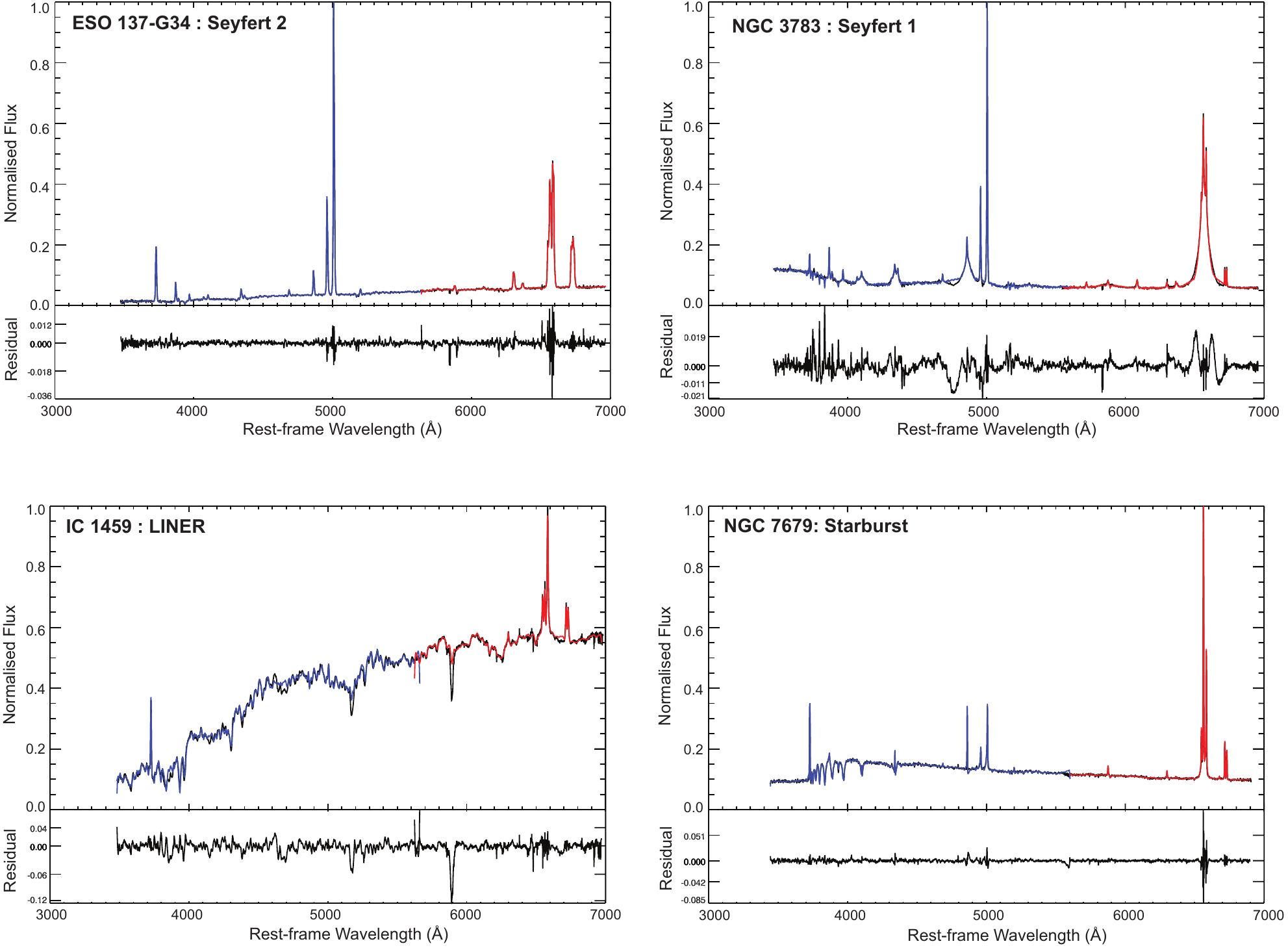}
\end{centering}
\caption{Representative fits to the nuclear spectra from S7. The instrumental resolution is comparable with the line thickness in these panels. We show one example of each of the major classes of object in the survey. Top left, ESO~137-G34, a typical Seyfert 2. Top right, NGC~3783, a typical Seyfert 1. The broad lines are not used when extracting the narrow-line spectrum from the fit. Bottom left, IC 1459 a typical LINER. The spectrum of the old stellar population is particularly prominent, as is the interstellar absorption in the Na D lines (not fitted). Bottom right, NGC~7679, a typical starburst or post-starbust (given the prominence of the underlying A-type stellar absorption). In this nuclear spectrum, the \4959_5007 lines are broader than the others, and we infer the presence of a very weak Seyfert 2 mixed with the \HII\ region spectrum.}\label{fig:fig2}
\end{figure*}

\begin{figure}[htb!]
\begin{centering}
\includegraphics[scale=0.85]{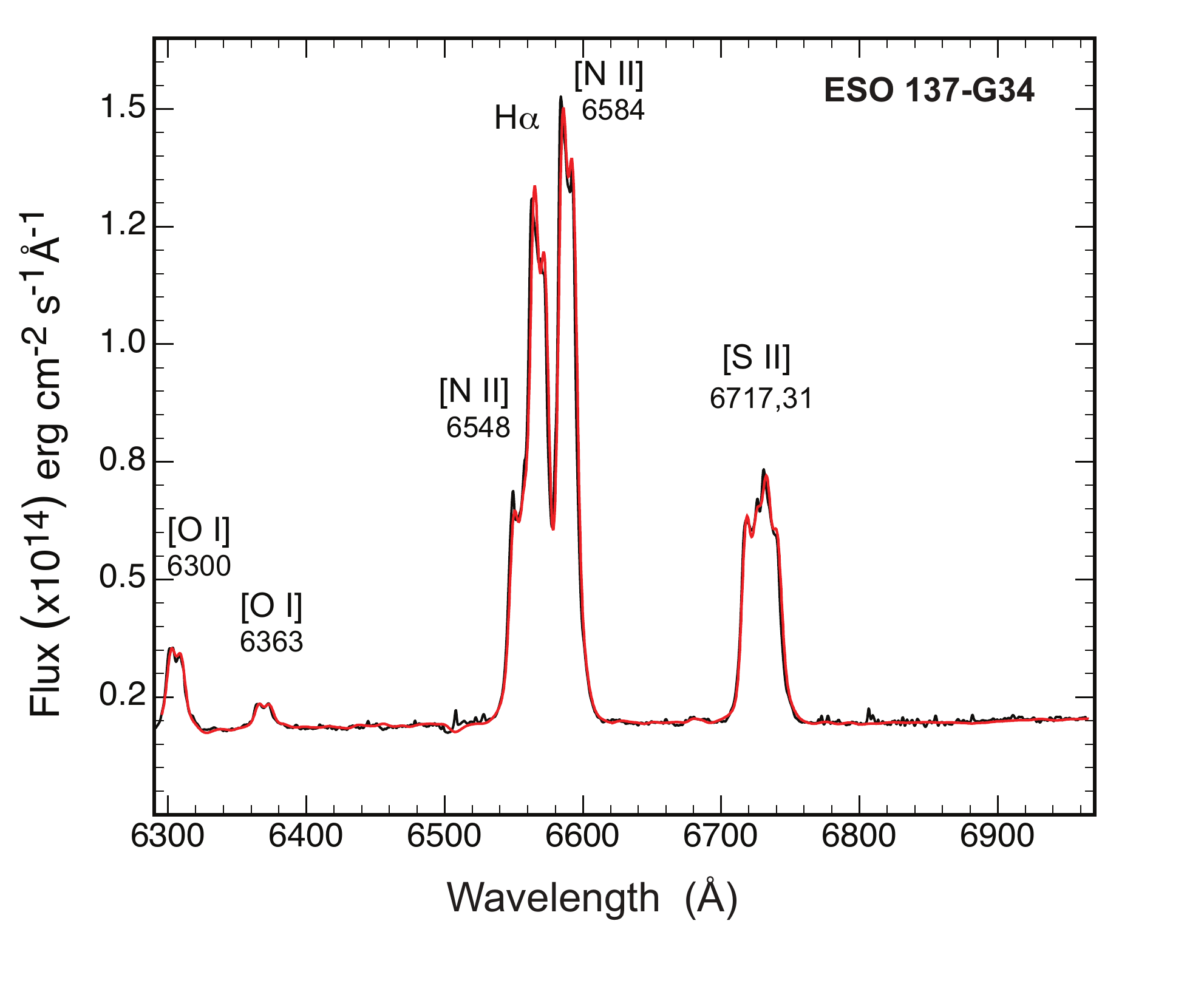}
\end{centering}
\caption{The fit (red line) to the observation (black line) for the Seyfert 2 ESO~137-G34 in the region of H$\alpha$. This demonstrates how the double-peak nature of the emission lines in this object is very well fit by the \lzifu\ code.}\label{fig:fig3}
\end{figure}

In Figure \ref{fig:fig2} we show the quality of the fits achieved in typical spectra representing each of the four major classes of object covered by S7; a Seyfert 1, a Seyfert 2, a LINER galaxy and a Starburst galaxy. This figure shows how both young and old stellar populations can be well fit. {In addition, double-line profiles can be reproduced, shown in the case of ESO~137-G34 in Figure \ref{fig:fig3}. In the case of the Seyfert 1 broad lines, the broad lines are not particularly well fit, but these are used only so as to allow a measurement of the NLR spectrum in these objects. The residuals on the fit are typically only of order of $1-2$\%, except in the case of the NaD absorption lines, which are not modelled.}

In order to determine the flux and extinction in the narrow-line region, and to extract the narrow line spectrum, we have used only the two narrowest components of the 3-component fit.  The justification for this is that the broad component was used only in the case of the recombination lines of the Seyfert 1 broad line region (see, for example, Figure \ref{fig:fig2}, NGC~3783). These arise from the nuclear BLR, rather than in the NLR, the spectrum and emission line ratios of which we seek to extract.

In Table \ref{table:Flux} we give the luminosity distance, extinction, size of the effective aperture at the distance of the object, the observed H$\beta$ and \lOIII\ fluxes, the inferred nuclear luminosities in these two emission lines, and the approximate size of the ENLR based upon the region of the data cube in which the \lOIII\ line is detected with a signal to noise greater than 5. In Table \ref{Velocity} , we list the velocity widths (FWHM) of the three fitted components, and in Table \ref{Lines} we give the measured emission line fluxes of all our objects, sorted according to their right ascension. 

\subsection{Seyferts with Coronal Lines} \label{coronal}
\begin{figure}
\centering
\includegraphics[width=0.75\columnwidth]{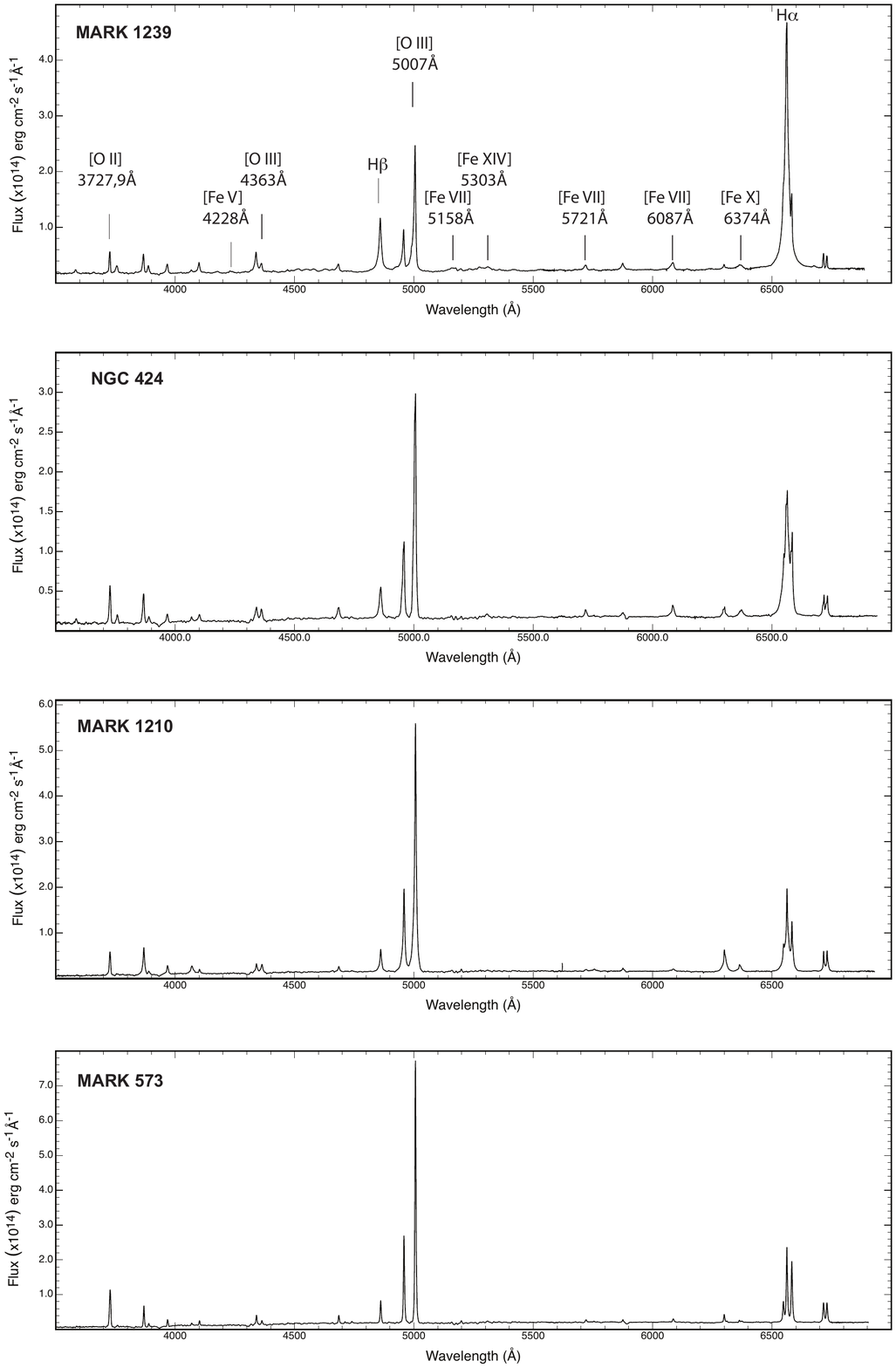} 
\caption{A selection of the S7 galaxy nuclear spectra showing strong coronal line emission. The main coronal species and a number of the other lines are identified on the first panel. The panels are ordered in terms of the electron density as indicated by the [O III] $\lambda \lambda 4363/5007$\AA\ ratio. MARK 1239 would be the densest with the electron density in the \OIII\ zone exceeding $10^6$cm$^{-3}$. Note that this ordering matches the order of the H$\alpha$ line width and of the [N II] $\lambda 6584$/H$\alpha$ ratio. This suggests the existence of correlations between the region of the narrow line emission, electron density, and the strength of the coronal features. }\label{fig:fig4}
\end{figure}

In our sample, and as noted in Table \ref{table:Obs}, we find a number of objects of both the Seyfert 1 and Seyfert 2 types which have coronal emission from species such as [\ion{Fe}{5}], [\ion{Fe}{7}], [\ion{Fe}{10}] and [\ion{Fe}{14}] in their spectra. In Figure \ref{fig:fig4} we show a number of these, representative of the Seyfert 2 objects that display coronal emission. It is evident from the figure that the relative strength and excitation of the coronal emission is correlated with the H$\alpha$ line width, and with the electron density in the \OIII\ -- emitting region, as evidenced by the \OIII\ $\lambda\lambda 4363/5007$ ratio. The density in the  \OIII\ zone must exceed $10^6$cm$^{-3}$ in the densest example (MARK 1239) while the low-excitation gas as revealed by the \SII\ $\lambda\lambda 6717/6731$ ratio remains around $n_e \sim 10^4$cm$^{-3}$, and the strength of these \SII\ lines and the \NII\ $\lambda 6584$ line relative to the broad component of H$\alpha$ also changes systematically with H$\alpha$ line width. These low excitation lines clearly arise in a region which is physically distinct from the region which emits the coronal species.

\begin{figure}
\centering
\includegraphics[scale=0.5]{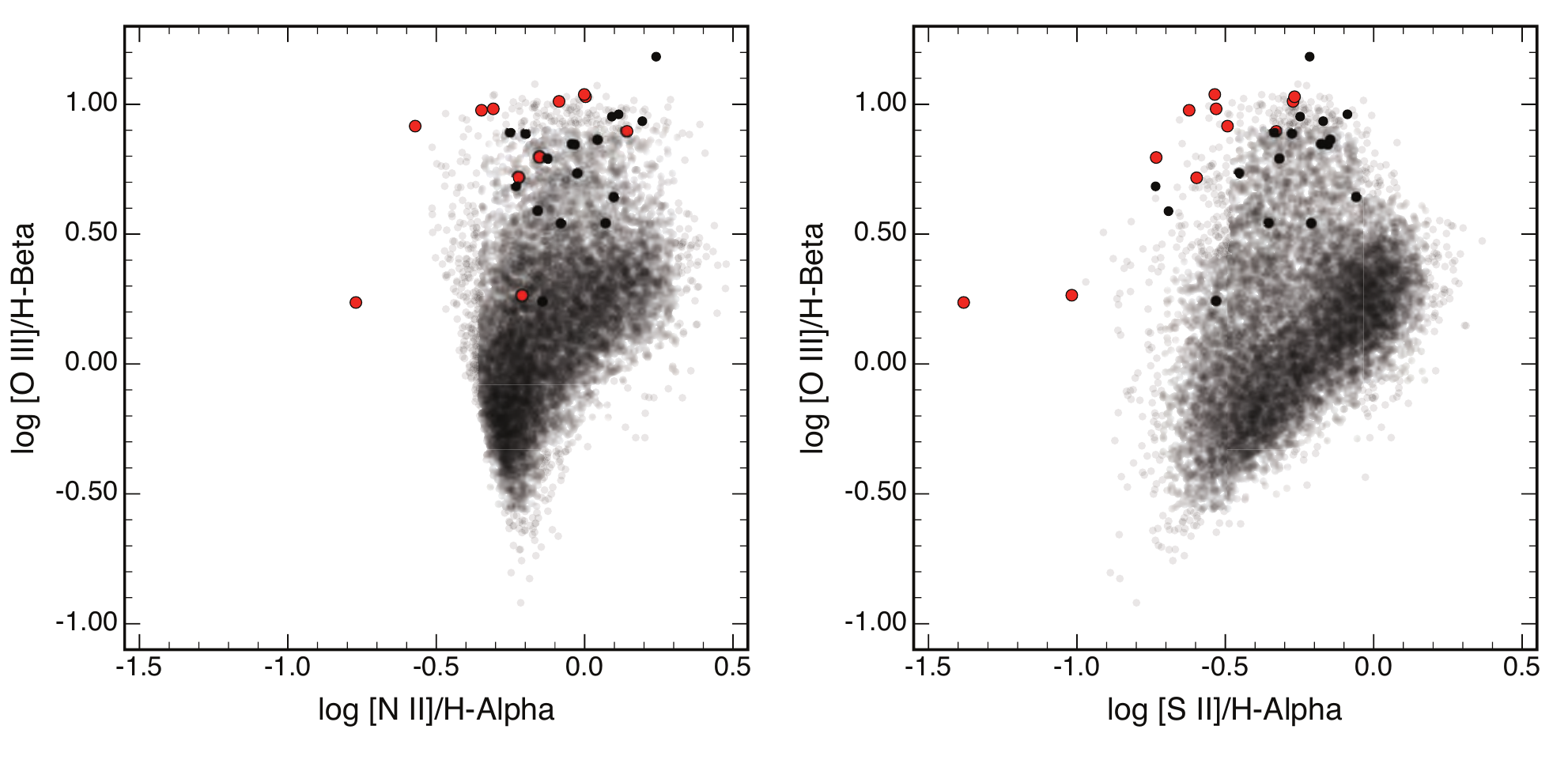} 
\caption{The \citet{Baldwin81} and \citet{Veilleux87} diagnostic plots for the SDSS AGN as isolated by \citet{Vogt14} (gray points), the Seyfert 2 galaxies as measured here (black points) and the Seyfert galaxies of any class from the S7 survey which display coronal emission (red circles). {The Seyfert galaxies in the S7 cluster near the upper limit of the SDSS points, which is presumably an aperture effect. Also the emission line ratios of the Seyfert galaxies with coronal emission displaced to the left relative to those without.} }\label{fig:fig5}
\end{figure}

A confirmation that the Seyferts with coronal lines form a distinct population is obtained by an examination of the \citet{Baldwin81} (BPT) and \citet{Veilleux87} (V\&O) diagnostic plots. In Figure \ref{fig:fig5} we plot the objects with coronal lines (red circles) and the S7 Seyfert 2s (black circles) against the AGN from SDSS as isolated by \citet{Vogt14}. From this Figure, it is evident that the objects with coronal lines {form a sequence which is somewhat separated from those without. } While normal Seyfert 2s are clustered in the upper part of the Seyfert branch as defined by the SDSS galaxies, {most} objects with coronal lines lie above and to the left. In general they have {\bf lower} [\ion{S}{2}]/H$\alpha$ ratios, and the more extreme objects also have lower [\ion{N}{2}]/H$\alpha$ and [\ion{O}{3}]/H$\alpha$ ratios. This is most easily understood as an effect of density, since the forbidden line ratios appear {to decrease in the order of their critical density. For example, [\ion{S}{2}] /H$\alpha$ decreases before [\ion{O}{3}]/H$\beta$ does, which would go some way towards explaining the distribution the coronal Seyferts on Figure \ref{fig:fig4}}.  From an inspection of the line widths, it is also clear that these are also ordered according to their critical density, with  [\ion{S}{2}] $\lambda 6717,31$ and \NII\ $\lambda 6584$ being the narrowest, \lOIII\ and [\ion{O}{1}] $\lambda 6300$ being broader, the coronal lines broader still and H$\alpha$ the widest. However, from these observations  it is not possible to dis-entangle the effects of rotation and outflow in determining the width of the various emission lines.

These properties are consistent with the model advocated by \citet{Mullaney09}. In this, the coronal lines arise in a dense gas which is launched from the dusty inner torus ($10^{17} < R/{\rm cm} < 10^{18}$) at very high local ionisation parameter $\log U \sim -0.4$, and which is accelerated by radiation pressure to a terminal velocity of a few hundred km~s$^{-1}$. At this ionisation parameter the gas is Compton heated to $\sim 10^6$K or greater and the dust in the coronal emission region is destroyed, allowing the forbidden iron lines to reach such high intensity relative to the hydrogen lines. Systematic changes in width, excitation and density are consistent with different inner torus radii for the objects shown in Figure \ref{fig:fig4} - MARK~1239 having a small inner torus, and MARK~573 having a relatively larger one. 

The fact that all our Seyferts lie in the upper part of the Seyfert branch defined by the SDSS galaxies suggests that the``band" of points which defines the Seyfert sequence on the BPT diagrams is in fact a mixing sequence (presumably  a result of aperture effects in the SDSS survey) between ENLR and \HII\ regions in the SDSS aperture. Such mixing has been investigated in individual galaxies by  \citet{Scharwachter11},  \citet{Dopita14}, and Davies et al. (2014a,b). The implication of this is that both the Seyfert nuclei and the \HII\ regions associated with them are all of super-solar metallicity.

\subsection{The ENLR and the circum-nuclear \HII\ regions}\label{Imageresults}
\begin{figure*}[htb!]
\begin{centering}
\includegraphics[scale=0.8]{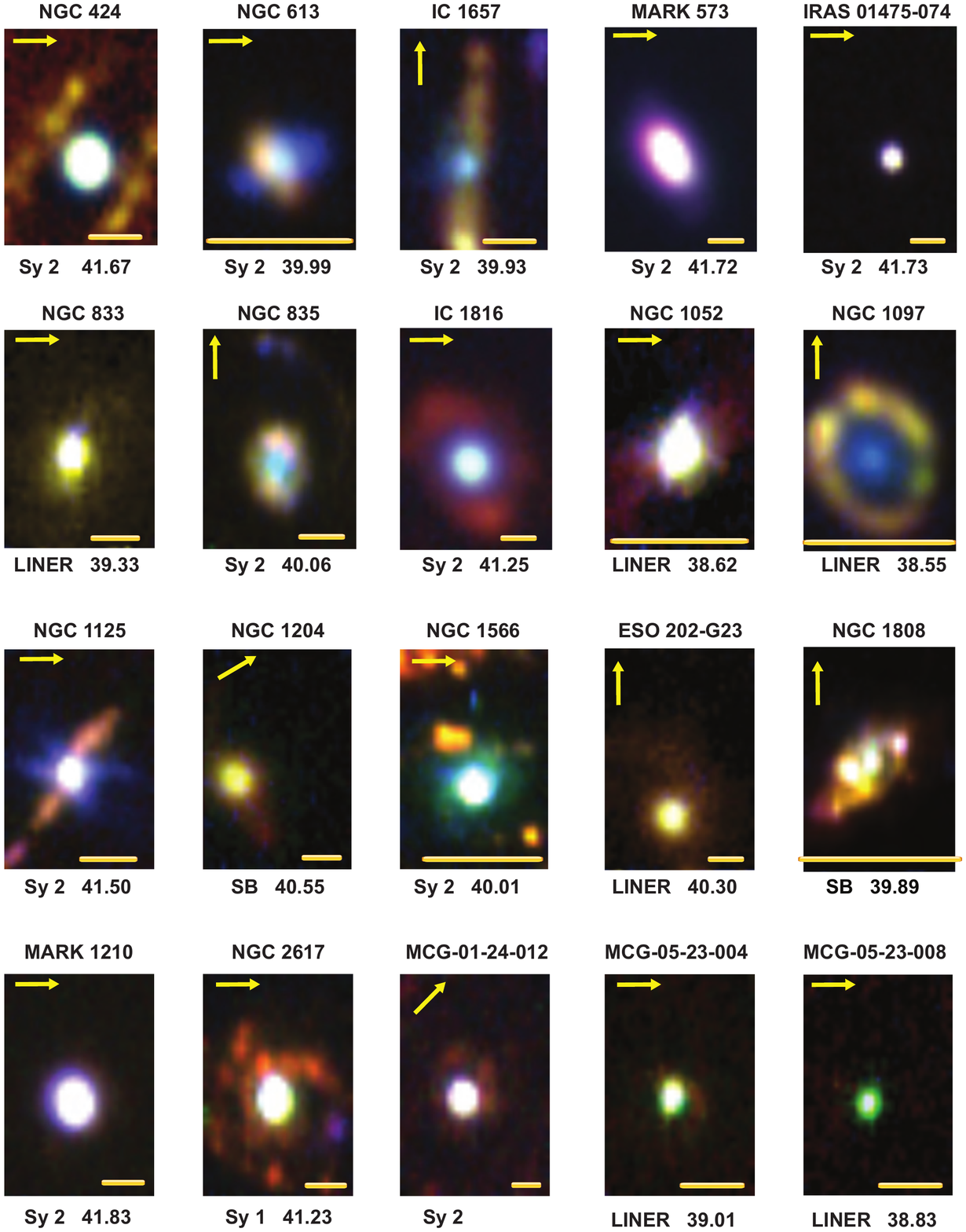}
\end{centering}
\caption{Color images of the galaxies lying between RA = 00Hr and 09Hr 44m. All images are $25\times37$~ arcsec in size. The direction of north is indicated by a yellow arrow. The galaxy type and nuclear \lOIII\ luminosity is given below each images. In these images H$\alpha$ is in the red channel, \lNII\ is in the green, and \lOIII\ in the blue channel, to distinguish the \HII\ regions (gold,yellow or red) from the Seyfert and LINER nuclei and their ENLR (blue, green, turquoise or violet).\newline}\bigskip \label{Images_1}
\end{figure*}

\begin{figure*}[htb!]
\begin{centering}
\includegraphics[scale=0.8]{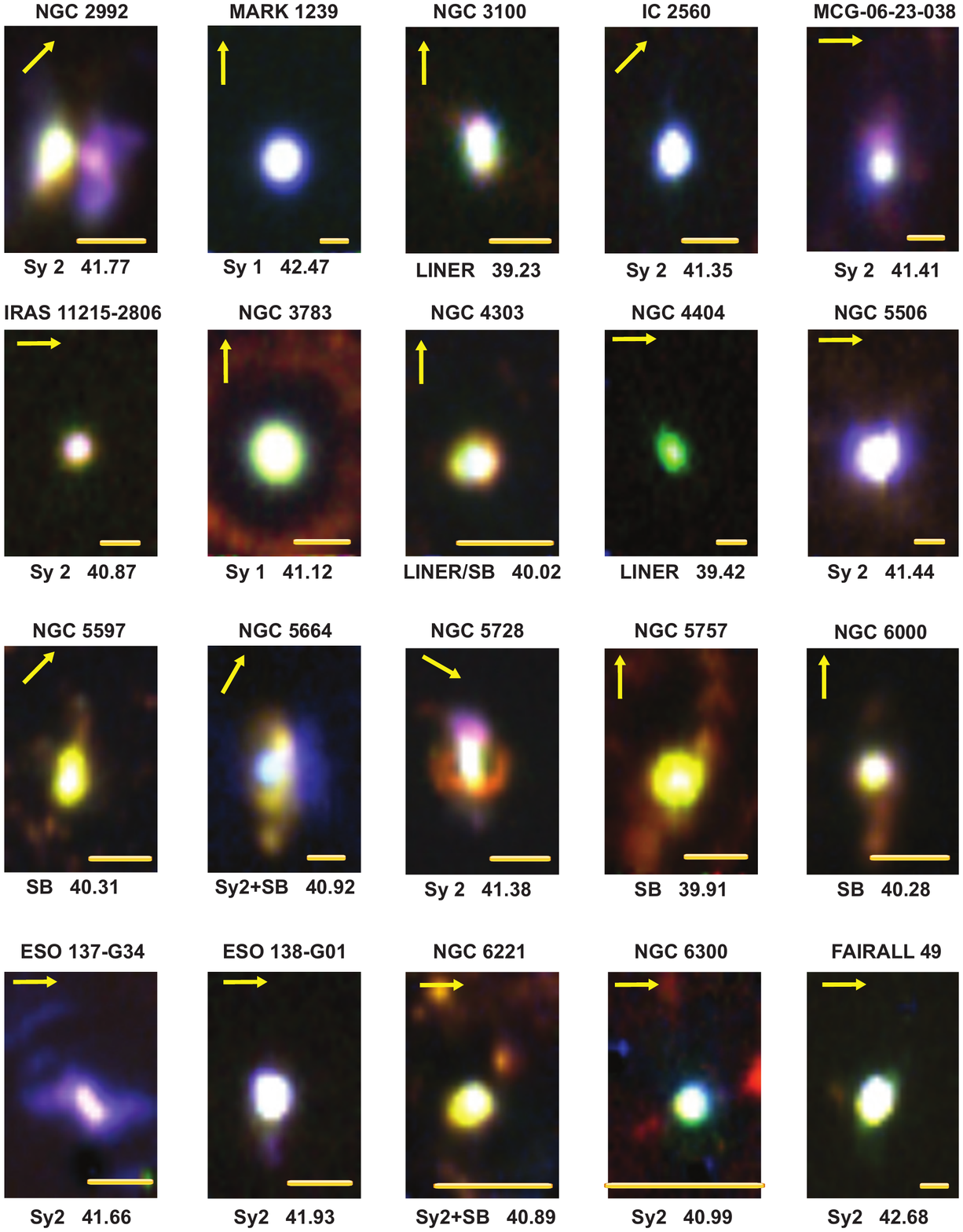}
\end{centering}
\caption{As Figure \ref{Images_1}, but for the galaxies lying between RA = 09Hr 45m and 18Hr 37m.}\label{Images_2}
\end{figure*}

\begin{figure*}[htb!]
\begin{centering}
\includegraphics[scale=0.8]{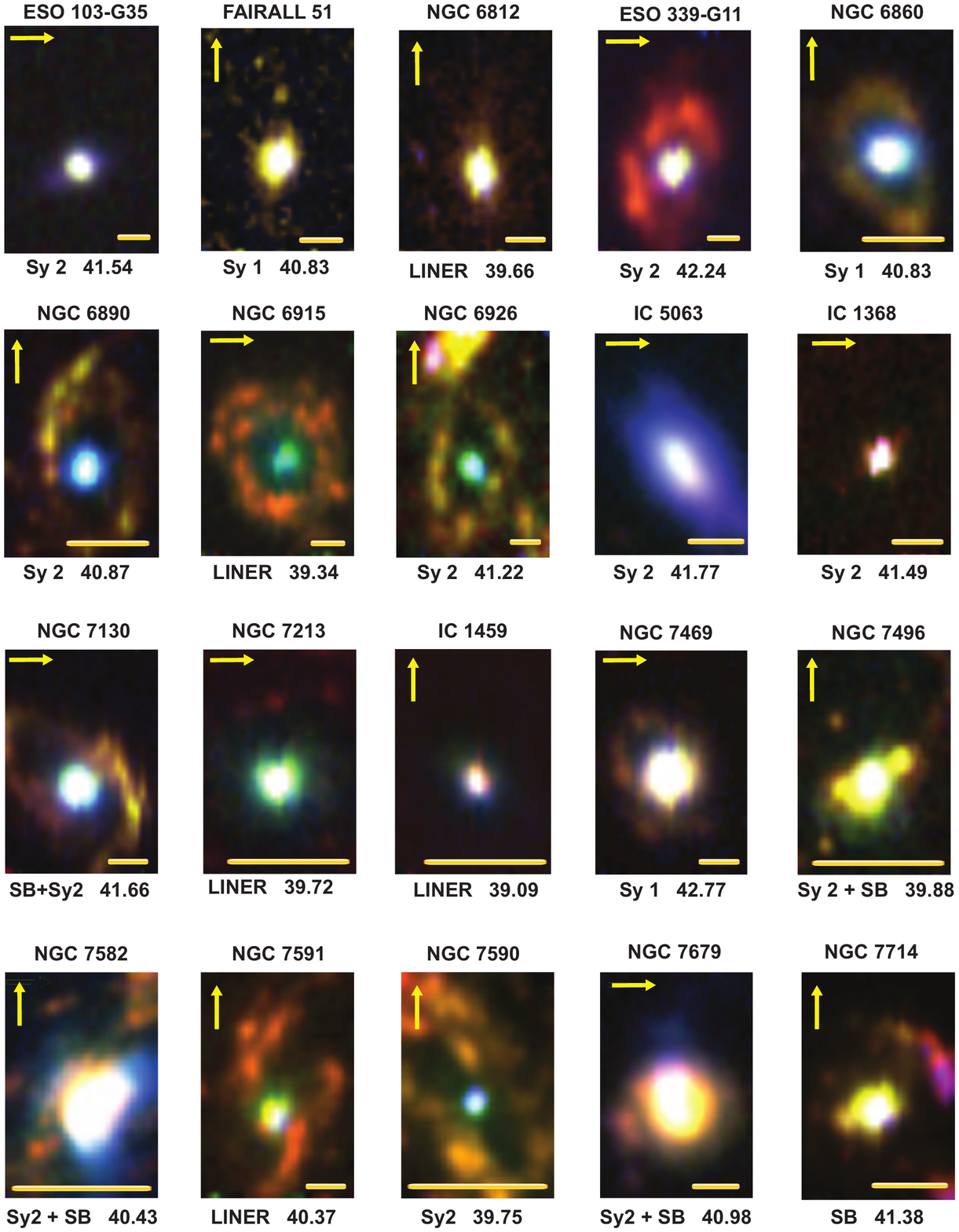}
\end{centering}
\caption{As Figures \ref{Images_1} and \ref{Images_2}, but for the galaxies lying between RA = 18Hr 38m and 24Hr.}\label{Images_3}
\end{figure*}

In Figures \ref{Images_1} to \ref{Images_3} we present colour images of the Seyfert galaxies in our survey. As explained above, the colour scheme is chosen so as to distinguish Seyferts and their ENLR from \HII\ regions. This uses the fact that the \HII\ regions are (at higher metallicities) dominated by H$\alpha$ while \lNII\ emission is strong, and  \lOIII\  is weak. Since H$\alpha$ is in the red channel and \lNII\ is in the green channel in these images, \HII\ regions appear as either red, orange or yellowish. In the Seyfert nuclei and their ENLR by contrast, \lOIII\ or \lNII\ are particularly strong. These regions therefore appear as violet, blue, turquoise green or even white.

{Each WiFeS image is $25\times37$arcsec in size; the final row of spaxels is noisy and has been removed. The orientation of the image is shown with a yellow arrow pointing N, and a scale bar 2 kpc long is given. The images are ordered in sequence of increasing RA.}


A few qualitative points can already be made on the basis of these images. First, we see that the conjunction of a  Seyfert ENLR lying inside a circum-nuclear ring or torus of star-formation (typically a few kpc across) is a fairly common phenomenon. It is clearly seen in 17 galaxies ($\sim 25$\%) of the sample. The star-forming ring often shows rapid rotation in the WiFeS data cube, and such features have been identified as an Inner Lindblad Resonance \citep{Wilson93}. We will investigate this point in a future paper in this series. Second, where the orientation of the disk can be established, the ``ionisation cones'' in these cases are arranged in the polar direction, perpendicular to the star-forming ring or torus, showing that the inner thick dusty torus of the accretion disk around the black hole must be essentially co-planar with the star-forming ring located further out. This geometry is seen clearly in 9 galaxies of our sample ($\sim 15$\%). Third, while all the lower-luminosity AGN are found to be associated with \HII\ regions lying within the WiFeS field, many of the higher luminosity objects do not display associated \HII\ regions, even though the WiFeS field in general covers a larger area of the galaxy. It is tempting to conclude that in these cases the flux of ionising photons is sufficiently high as to largely ionise the surrounding ISM and suppress star formation. This is an idea that has been developed by \citet{Curran13} in the effort to understand why \ion{H}{1} absorption is rare in the hosts of high-redshift AGN. These ideas will be more fully investigated in future papers using the full S7 sample.

\section{Notes on Individual Objects}\label{Notes}
\noindent
{\bf PKS 0056-572: } The object has been mis-classified in \citet{VC06}. We conclude that this object is a QSO at a redshift of z = 1.46 - see Figure \ref{fig:fig1}. \newline
{\bf NGC424: }	This Seyfert 2 galaxy has been imaged with HST by \citet{Malkan98}. It contains a bright ENLR with relatively broad and strong \ion{He}{2} line emission detected out to $\sim10$ arcsec (2.2~kpc) diameter. A Seyfert 1 nucleus is weakly detected at H$\alpha$, with prominent coronal line emission - see Figure \ref{fig:fig4}. Broad H$\alpha$ and H$\beta$ has been detected in polarised light from the nucleus by \citet{Moran00} with a width of 12000 km~s$^{-1}$.  The nucleus is flanked by two sets of \HII\ regions within a fast-rotating larger star-forming ring.\newline
{\bf NGC 613: }	This galaxy contains a very fine bipolar ENLR $\sim 25\times 10$ arcsec ($\sim 2 \times 0.8$kpc). The two lobes of emission are poorly aligned with the radio lobes detected by \citet{Condon87}. The ENLR lobes emerge from a disk of  \HII\ regions, strongly dominant close to the nucleus.\newline
{\bf IC 1657: } Displays a very weak fan-like Seyfert 2 ENLR emerging from one side of a nearly edge-on galaxy disk with many \HII\ regions. The ENLR ionisation cone with an opening angle close to 90 $\degr$ is visible extending $\sim 11$ arcsec (2.4~kpc) from the plane of the galaxy out to the edge of the field. The nucleus displays a strong continuum of old stars, and the emission spectrum is dominated by \HII\ regions rather than by the ENLR. \newline
{\bf MARK 573: }  This Seyfert 2 galaxy ENLR has been imaged in [\ion{O}{3}] by \citet{Schmitt03b}, revealing a complex knotty ionisation cone on a $\sim 10$ arcsec ($\sim 2.2$kpc) scale. The WiFes data shows  a very bright and much more extensive barrel-shaped ENLR which is over 8~kpc long. There is no sign of embedded \HII\ regions. The nuclear spectrum shows strong coronal emission, see Figure \ref{fig:fig4}.  \citet{Fischer10} using Wide Field Planetary Camera 2 and STIS observations argue that the ionising radiation field, confined to an ionisation cone from the central engine is inclined and intersects and ionises the inner spiral arms of the galaxy  driving these into outflow. \newline
{\bf IRAS 01475-0740: }  The nucleus of this Seyfert 2 galaxy is very bright, but the ENLR can be traced in the WiFeS data cube over some $\sim 10\times 7$ arcsec ($\sim 3.3 \times 2.3$kpc). The HST [\ion{O}{3}] image by \citet{Schmitt03a} is dominated by the compact nucleus. Broad lines have been detected in polarised emission \citep{Tran03}, showing that the AGN core of the galaxy is heavily obscured. \newline
{\bf NGC 833: } This LINER galaxy is tidally interacting with NGC 835 and these are two members of the interesting and highly active Hickson Compact Group 16. All galaxies in this group are either active or starburst. The equivalent width of the [\ion{N}{2}] lines are quite high, and the [\ion{O}{3}] and [\ion{O}{1}] lines are quite strong, superimposed on a strong old star continuum.\newline
{\bf NGC 835: } This Seyfert 2 galaxy is tidally interacting with NGC 833, and dispelays a long tidal tail. The nuclear spectrum is dominated by a strong post-starburst continuum with deep NaD absorption.  In X-rays, \citet{Turner01} confirmed the presence of an AGN in both NGC 835 and NGC 833. The ENLR, $\sim 8\times 5$ arcsec ($\sim 1.9 \times 1.3$kpc), is contained within an elliptical starburst ring $\sim 14\times9$ arcsec ($\sim 3.4 \times 2.3$kpc) in diameter.\newline
{\bf IC 1816: } This Seyfert 2 galaxy with a bar + ring structure and tidally disturbed outer arms \citep{Malkan98}. The ENLR has been studied by \citet{Fehmers94} and  \citet{CidFernandes98}, and long slit spectroscopy obtained by \citet{Fraquelli03}. The WiFeS nuclear spectrum displays fairly broad forbidden lines and weak coronal lines - [\ion{Fe}{7}] 6087, 5721 \AA\ lines are visible.  The ENLR fills the region within the prominent star-forming ring located as a radial distance of $\sim 3$kpc.\newline
{\bf NGC 1052: 	} This classical liner shows an X-shaped cross of strong [\ion{O}{3}]  emission, and an extended ridge of H$\alpha$ emission. A detailed accretion plus jet cocoon shock model for the excitation of this complex source is presented in the next paper of this series (Dopita et al. 2015, in press) \newline
{\bf NGC1097: } A tidally-distorted grant-design spiral with a weak active nucleus embedded in bright stellar continuum, surrounded by a star-forming ring with E+A spectrum. Broad NaD absorption is evident around the nucleus. The nucleus displays a high velocity dispersion with very strong \NII. A faint ENLR re-appears outside the star-forming ring.  A study by \citet{Fathi06} using the GMOS-IFU as well as high resolution HST-ACS observations shows that the nuclear activity appears to be fed by radial flow from the star-forming ring towards the nucleus. The nuclear activity is variable, and the galaxy has been classified as a Seyfert~1 on the basis of broad H$\alpha$ when more active \citep{Storchi-Bergmann97,Storchi-Bergmann03}.\newline
{\bf NGC 1125:	} This Seyfert 2 galaxy contains a superb ionisation cone extending not quite at right angles from the disk which is seen nearly edge on. The ENLR appears as a double paraboloidal structure, and is dynamically quiescent, co-rotating with the disk gas. The orientation of the ENLR is well aligned with the much larger $\sim 12$kpc double-lobe structure seen at 8.4GHz \citep{Thean00}. There are a number of bright \HII\ regions in the disk. An HST F606W continuum image has been published by  \citep{Malkan98}. \citet{Mulchaey96a} finds that the  [\ion{O}{3}] emission is only slightly resolved with HST - presumably due to surface brightness limitations. \newline
{\bf NGC 1204:	} Appears to be a fairly metal-rich starburst. The nuclear spectrum shows a strong, heavily reddened late- or post- starburst continuum. 
\citet{Zaw09} note that the line ratios in this object put it in the \citet{Kewley06} transition zone, suggesting that shocks may be important in powering this source. A broad component to H$\alpha$ is detected at the nucleus.\newline
{\bf NGC 1566:	} A superb two-arm spiral with very active star formation, this Seyfert~1 galaxy has an ENLR $\sim 15\times 10$ arcsec ($\sim 1.5 \times 1.0$kpc) extended in the NE-SW direction. The circum-nuclear reddening is high, so the ENLR is more easily traced in \NII\. A broad component to H$\alpha$ is detected at the nucleus, confirming the Seyfert~1 type given on the basis of HST FOS observations \citep{Kriss91} .\newline
{\bf ESO 103-G35: } The nuclear emission line spectrum suggests that this object is a very heavily reddened Seyfert 2, or possibly a LINER with vey strong \NII\ and [\ion{O}{1}] . The underlying continuum is also very red. The disturbed morphology of the galaxy suggests that it is a post-merger.\newline
{\bf NGC 1808: } This galaxy is simply a circum-nuclear ($\sim 1$kpc) starburst galaxy with very deep NaD absorption, and a late-starburst continuum spectrum. It is hard to understand why it was ever assigned a Seyfert~2 class, but the \citet{VC06} catalog now has it as H2 type.\newline
{\bf MARK 1210: }  This galaxy has been imaged with HST in the F606W filter by \citet{Malkan98}. All the emission lines in this Seyfert~2 galaxy have a broad ($\sim 475$km~s$^{-1}$) component. The ENLR is confined within the central $\sim 1.0$kpc, in agreement with the IR results of \citet{Mazzalay07}. The coronal lines are strong - see Figure \ref{fig:fig4}, and the IR coronal spectrum has been extensively studied by \citet{Mazzalay07}. Coronal lines of [\ion{S}{8}], [\ion{S}{9}], [\ion{Si}{7}], [\ion{Si}{10}],[ and [\ion{Ca}{8}] were  detected.\newline
{\bf NGC 2617: } This Seyfert~1 galaxy displays extremely broad Balmer line emission with a line profile that is steeper on the blue side. By contrast the ENLR lines are relatively narrow ($\sim 120$km~s$^{-1}$). The ENLR is elliptical  $\sim 9\times 6$ arcsec ($\sim 2.6 \times 1.7$kpc) and orientated almost E-W, but here is a prominent ENLR / \HII\ region knot at PA$ = 330\degr$ and at 15 arcsec from the nucleus ($\sim 4.4$kpc). The nucleus is surrounded by many bright \HII\ regions.\newline
{\bf MCG-01-24-012: } This is a Seyfert 2 galaxy detected in hard X-rays by BeppoSAX by \citet{Malizia02}. It was imaged using HST by \citet{Schmitt03a}, who found the ENLR to be extended by $880 \times 440$pc at PA$=750\degr$. In the WiFeS data cube, the ENLR can be traced further, out to  $3.7 \times 2.5$kpc, with a major axis agreeing with that measured with HST. The ENLR is surrounded by a faint ring of \HII\ regions about 3kpc in radius.\newline
{\bf MCG-05-23-004: } This poorly-studied late-type galaxy is a LINER with LINER-like spiral emission extensions. These can be traced out to $\sim 12$ arcsec (2.3kpc) from the nucleus. \newline
{\bf MCG-0105-23-008: } This is also a faint LINER in an Elliptical galaxy. \newline
{\bf NGC 2992: 	}This galaxy is a tidally-distorted galaxy interacting strongly with NGC~2993. The very extensive ENLR has been studied and interpreted through multi-slit spectroscopy by \citet{Allen99}. The galaxy displays an extensive biconical ENLR with large opening angle extending above and below a nearly edge-on dusty disk. The nucleus exhibits variability, with a Seyfert~1 nucleus appearing and fading, apparently corralated with X-ray activity. The original spectra by \citet{Ward80} showed broad wings to H$\alpha$, while in 1997 these had disappeared \citep{Allen99}. By 2000 these had returned \citep{Gilli00}, but then once again disappeared in the observations by \citet{Trippe08}, who argue against the hypothesis that these variations are simply due to variable extinction in the circum-nuclear dust lane seen with HST \citep{Malkan98}. In our spectrum, we once again see broad wings to H$\alpha$. These wings have been detected in polarised light \citep{Lumsden04}. \newline
{\bf MARK 1239: }  This E-S0 galaxy contains a very luminous Seyfert 1 with a very rich and high-excitation coronal spectrum including  [\ion{Fe}{14}] - see Figure \ref{fig:fig4}. The ENLR is very bright and extensive with an elliptical morphology ($\sim 17\times 15$ arcsec, $\sim 7.1 \times 6.3$kpc). There is no sign of any \HII\ regions. \newline
{\bf NGC 3100: 	} This is a LINER in a S0 galaxy with - in common with other LINERS - very low extinction, but with deep NaD absorption. \newline
{\bf IC 2560: } A Seyfert 2 galaxy with relatively narrow lines and coronal emission. There is a hint of a broad component to H$\alpha$. The ENLR is very extensive ($\sim 18\times 13$ arcsec, $\sim 3.8 \times 2.8$kpc) with a jet-like extension towards the N. \newline
{\bf MCG-06-23-038: } The ENLR  ($\sim 11\times 9$ arcsec, $\sim 3.5 \times 2.9$kpc)  is dynamically active showing pronounced line splitting, The ENLR is embedded in a rotating disk of \HII\ regions elongated along the major axis of the galaxy. There are a number of other \HII\ regions scattered across the field, associated with the outer stellar ring. \newline
{\bf IRAS 11215-2806: } This Seyfert 2 galaxy has a fairly compact ENLR  ($\sim 7\times 7$ arcsec, $\sim 2.1$kpc). It has been imaged with HST in the F606W filter by \citet{Malkan98}. \newline
{\bf NGC 3783: } This barred ring spiral has been imaged with HST in the continuum by \citet{Malkan98}, and the Seyfert1 nucleus in \ion{O}{3}] by \citet{Schmitt03a}. However, the [\ion{O}{3}]  is almost point-like in the HST image. The measured [\ion{O}{3}]  luminosity $\log{\rm L} = 41.08$ agrees closely with our measurement agrees closely with our measurement - $\log{\rm L} = 41.12$. We find this compact nucleus, strong in coronal line emission,  to be surrounded by an extensive low-surface brightness ENLR $\sim 18$ arcsec, or $\sim 3.8$kpc in diameter, effectively filling  the central region out to the star forming ring, marked by \HII\ regions.\newline
{\bf NGC 4303: } The nuclear spectrum is that of a late starburst, or possibly a LINER type. However, [\ion{O}{1}] $\lambda 6300$ is very weak for a LINER. The nucleus is surrounded by a star-forming ring  $\sim 9$ arcsec (1.1kpc) in diameter \citep{Mazzuca08}, and many \HII\ regions are scattered across the field -- c.f. H$\alpha$ image from \citet{Banfi93}. \newline
{\bf NGC 4404: } This is a very weak and strongly reddened LINER, apparently in interaction with NGC 4403. \newline
{\bf 3C 278: }  This object, VV~201, is a closely-interacting double early-type galaxy and a very strong radio source. No emission is visible at the position of the double-lobe radio source nucleus \citep{Morganti93}. The spectrum is simply that of an Elliptical galaxy. \newline
{\bf NGC 5506: }  This disk galaxy is seen nearly edge-on, and a strong lane of absorption crosses the nucleus on the F606W HST image of  \citet{Malkan98}. The WiFeS nuclear spectrum of this Seyfert~2 galaxy shows weak coronal emission. \citet{Nagar02} classify it as an obscured narrow-line Seyfert 1 on the basis of a near-IR spectrum.  The ENLR is elongated in the NE direction at high flux levels in agreement with the IR measurements by  \citet{Raban08}, but at low flux levels appears more a a double fan with wide opening angle orientated N-S at right angles to the stellar disk. Its full extent is $\sim 15\times 12$ arcsec or $\sim 2.1 \times 1.6$kpc. \newline
{\bf NGC 5597: }  This somewhat tidally-disturbed galaxy forms  an interacting physical pair with NGC 5595. The nuclear region forms a bar of star formation imaged  with HST in the red continuum by \citet{Malkan98}. The nuclear continuum spectrum is dominated by young stars, and a broad \ion{He}{2} $\lambda 4686$ feature from a WR population is faintly visible. A number of other \HII\ regions are visible in the general field. \newline
{\bf NGC 5664: } This poorly-studied Seyfert~2 is embedded in a star-forming ring $\sim 13$ arcsec (4.0kpc) in diameter. It displays a magnificent bi-cone ENLR orientated in the polar direction of this disk, with an opening angle of $90\degr$. This is much fainter on the east side. The ENLR extends right across the WiFeS FOV, and is therefore larger than $\sim 25$ arcsec (7.7kpc) across. The ENLR is co-rotating with the galaxy, and is dynamically quiescent. \newline
{\bf NGC 5728:	} A barred ring spiral galaxy with a Seyfert~2 nucleus, it displays a well-collimated bright two-sided ENLR first described by \citet{Wilson93} on he basis of HST images. The ENLR emerging SE-NW at right angles to a star forming ring with a diameter of $\sim 2$kpc. The length of the ENLR is $\sim 22$ arcsec (4.4kpc) across - much larger than the 1.8 kpc reported by \citet{Wilson93}. The inner ENLR has strong line splitting in \lOIII. The nuclear spectrum shows a Seyfert~2 spectrum superimposed on a continuum spectrum showing signs of recent star formation. The [\ion{Fe}{7}]$\lambda$ 6087 is present, but is very weak compared to the continuum. In the IR the coronal region has been detected in [\ion{Ca}{8}] \citep{Emsellem01} and  in its [\ion{Si}{6}] emission \citep{Riffel06}. \newline		
{\bf NGC 5757:	} This barred spiral galaxy has a nuclear starburst with low reddening and high metallicity \citep{Saraiva01} embedded in spiral arms crowded with \HII\ regions. It has been imaged with HST in the continuum by \citet{Malkan98}. \newline
{\bf NGC 6000:	} A nuclear starburst on a barred spiral, very similar to NGC~5757, but with deep NaD absorption. It has been imaged with WFPC2 in the F606W filter with HST by \citet{Carollo97}. \newline
{\bf ESO 137-G34:} A spectacular Seyfert 2 with a pronounced line splitting - see Figure \ref{fig:fig3} - and a very extensive bi-conical ENLR with an opening angle of $80\degr$ and containing multiple shells. The ENLR extends beyond the WiFeS field, and is therefore in excess of $ 40$ arcsec (7.4kpc) in diameter. The nucleus has been imaged with HST in the red continuum by \citet{Malkan98}, whilee the ENLR has been studied with HST by \citet{Mulchaey96a}.\newline
{\bf ESO 138-G01:}  A luminous Seyfert~2 with a rich coronal-line spectrum. There is a suggestion of a broad-line component underlying H$\alpha$. The ENLR of this  galaxy has been imaged by \citet{Schmitt05}, who traced the ENLR over about 16 arcsec, out to the inner jet region visible in Figure \ref{Images_2}. However, in the WiFeS data cube, we see the ENLR extending over the full field, out to at least $ 35$ arcsec (6.4kpc) in diameter. In particular, there are two knots of emission at $\sim 15$ arcsec either side of the nucleus at PA $= 105\degr$. Both show pronounced line splitting with $\Delta v \sim 350$km~s$^{-1}$. \newline
{\bf NGC 6221: } This galaxy, imaged in H$\alpha$ by \citet{Ryder93} and in the optical and IR at high resolution by \citet{Levenson01} and also studied spectroscopically by \citet{Gu06} contains a very weak Seyfert~2 nucleus (detected by the broader \lOIII\ line) embedded in a bright starburst nucleus. This galaxy is an X-ray-loud composite galaxy, with very weak AGN activity seen at optical or IR wavelengths , but displaying strong X-ray emission. Obscuration by a column of $N_{\mathrm H} = 10^{22}$cm$^{2}$ is sufficient to explain these properties \citep{Levenson01}. Many \HII\ region complexes are scattered across the WiFeS field of view. \newline
{\bf NGC 6300: } This barred ring spiral galaxy has a heavily obscured Seyfert~2 nucleus with deep NaD absorption. Long slit spectroscopy obtained by \citet{Fraquelli03}. This galaxy changes from Compton-thick to Compton-thin at X-ray wavelengths over timescales of a few years \citep{Matt03}, and contains an H$_2$O maser \citep{Greenhill03}. The ENLR is approximately $7\times 5$ arcsec, or $0.5 \times0.3$kpc in diameter in \lOIII\ but can be traced further out in the NE-SW axis in \lNII. A number of \HII\ region complexes are scattered across the WiFeS field of view. \newline
{\bf FAIRALL 49:} This ultra luminous IR galaxy with a compact nucleus  \citep{Malkan98} is a reddened Seyfert 2 with fairly broad forbidden lines and weak coronal lines of [\ion{Fe}{7}] 6087, 5721\AA. The ENLR can be traced across much of the field ($\sim 28\times 10$ arcsec, $\sim 11 \times 4$kpc). Broad H$\alpha$ and H$\beta$ components are detected in the nuclear spectrum. These components have also been detected in polarised light \citep{Lumsden04}. \newline
{\bf ESO 103-G35:} A Seyfert~2 galaxy studied by \citet{Gu06} with an extensive clumpy bipolar ENLR with an opening angle of $\sim45\degr$ orientated at PA $= 25\degr$ (roughly in the polar direction of the S0 host galaxy) and extending across the field of view of WiFeS ($27$ arcsec (10kpc). Away from the nucleus the ENLR lines are narrow and dynamically inactive.\newline
{\bf FAIRALL 51:} A Seyfert 1 with broad-line components visible in \ion{He}{1}, \ion{He}{2}, and down to H$\delta$.  The nuclear spectrum displays a strong high-excitation intermediate-width coronal spectrum with [\ion{Fe}{14}] 5303\AA\ and  [\ion{Fe}{10}] 6374\AA\ lines prominent. Other forbidden lines are narrower. The continuum of this galaxy is highly polarised \citep{Smith04}. The ENLR can be traced over $\sim 10\times 7$ arcsec, $\sim 2.8 \times 1.9$kpc. A few faint \HII\ regions are also visible in the field.\newline
{\bf NGC 6812:	} A classical LINER galaxy with E-type continuum and deep NaD absorption. This galaxy is in a group of 3, and may be interacting. The emission line region is elongated in the N-S direction (the polar direction in the galaxy), has dimensions  $\sim 24\times 13$ arcsec, $\sim 7.2 \times 3.9$kpc. Its velocity structure suggests that it is probably an outflow. \newline
{\bf ESO 339-G11:} A luminous IR Sb galaxy with a Seyfert 2 nucleus displaying strong \NII\ and weak coronal lines on reddened, late-type stellar spectrum. The $\sim 9\times 5$ arcsec ($\sim 3.3 \times 1.8$kpc) ENLR is located within a $\sim 3.5$kpc radius rapidly rotating starburst ring of \HII\ regions. \newline
{\bf NGC 6860:	} The spiral host galaxy is classified as a LIRG.  The Seyfert 1 nucleus is surrounded by ENLR $\sim 15$ arcsec  (4.3kpc) in diameter surrounded by a distorted ring of off-nuclear \HII\ regions with A-type absorption lines. This galaxy has been studied studied by high spatial resolution optical imaging and optical and near-IR spectroscopy by \citet{Lipari93}. The inner ENLR has been imaged in \lOIII\ by \citet{Schmitt03a}. \newline
{\bf NGC 6890: } This SA ringed spiral has a nucleus of  Seyfert~2 type with coronal emission. Long slit spectroscopy has been obtained by \citet{Fraquelli03}, and the high-excitation mid-IR lines have been investigated by \citet{Pereira-Santaella10}. It contains a bright region of broad \lOIII\ in an ENLR $\sim8\times8$ arcsec (1.2kpc), inside a bright and complex elliptical star-forming ring with major axis $\sim 28$ arcsec (4.2kpc) with many bright \HII\ regions.\newline	
{\bf NGC 6915:	} This is a poorly-studied LINER embedded in a double ring of \HII\ regions. The LINER region lies within the ring, and shows most extension in the polar direction. The velocity structure suggests that it is in outflow in the NE direction.\newline
{\bf NGC 6926:	} A Seyfert 2 nucleus with a dynamically-disturbed ENLR. This source contains an H$_2$O megamaser \citep{Greenhill03}. Many bright HII regions are seen in what appears to be a tidally-distorted LIRG disk galaxy. A narrow-band H$\alpha$ image has been published by \citet{Dopita02b}. The region to the north of the nucleus may contain a low-metallicity \HII\ region complex, based on its weak \lNII\ and strong  \lOIII\ relative to H$\alpha$. We may speculate that this galaxy is undergoing a minor merger.\newline
{\bf IC 5063: }  A post-merger Elliptical, this galaxy containing a Seyfert 2 nucleus, it shows a spectacular, symmetric ENLR extending across the WiFeS field. Barrel-shaped with a bright linear core in the inner in the inner regions, it develops horns with an opening angle of $\sim30\degr$ in the outer parts. Observed with HST, the inner region forms a fine knotty ionisation cone in \lOIII\ with an opening angle the same as in the outer regions\citep{Schmitt03a}. A small (4~arcsec; 860pc) double-lobe  plus nucleus radio structure has been resolved at 8.6GHz by \citet{Morganti98}, who also detected a very broad outflowing ($\sim 700$km~s$^{-1}$) HI absorption against the strong nuclear continuum source. The broad-line nucleus was detected in polarised light by \citet{Lumsden04}. No \HII\ regions are visible.\newline
{\bf IC 1368: } Seyfert 2 on old stellar continuum. The ENLR is very compact, $\lesssim 5$ arcsec and elongated in PA$=135\degr$. A few faint \HII\ regions are visible in the stellar disk, which is nearly edge-on at  PA$=40\degr$. These \HII\ regions are what causes the extension in the H$\alpha$ + \NII\ image of \citet{Colbert96}.\newline
{\bf NGC 7130 / IC5135: } In agreement with the classification of \citet{Thuan84}, the WiFeS nuclear spectrum shows this as  mixed excitation starburst + Seyfert 2. The \lOIII\ lines are appreciably broader than the H$\beta$, and there is a strong continuum from young stars mixed with the older stellar component.Long slit spectroscopy has been obtained by \citet{Fraquelli03}. This galaxy contains ENLR $\sim9\times9$ arcsec (2.7kpc) within a bright tidally-distorted barred spiral of \HII\ regions, nicely imaged with HST \citep{Malkan98}. \newline
{\bf NGC 7213: } This large SA galaxy contains a very bright LINER or Seyfert 1 nucleus with broad H$\alpha$. An old stellar continuum is prominent in the WiFeS spectrum. Broad \lNII\ can be traced over $\sim9\times9$ arcsec (0.9kpc). A number of \HII\ regions are visible towards the edge of the field - see also \citet{Evans96}. Recently \citet{Schnorr-Muller14} have presented high spatial resolution ($\sim 60$pc) integral field spectroscopy of the inner $0.8\times1.1$kpc of this galaxy using the Gemini Multi-Object spectrograph, and have inferred a black hole mass of $\sim 10^8$M$_{\odot}$, and a mass inflow of a  few tenths of a solar mass per year.\newline
{\bf IC 1459: } A LINER with deep NaD absorption, and with extended diffuse emission, which can be traced to a radial distance of over $20$~arcsec ($>2kpc$) towards the SW in  \lNII\  and H$\alpha$. The nucleus host a GPS radio source \citep{Tingay03}, and this source may have features in common with NGC1052 and NGC 7213. \newline
{\bf NGC 7469:	} A luminous Seyfert 1 nucleus with coronal emission and a fairly strong and extensive ENLR. A number of \HII\ regions form an irregular star forming elliptical ring approximately 20 arc sec in diameter (6kpc). The ENLR is  $\sim16\times8$ arcsec ($4.8\times2.4$kpc) in diameter.\newline
{\bf NGC 7496: } A very weak Seyfert 2 + starburst nucleus. Like NGC 7130, the galaxy nuclear spectrum shows  \lOIII\ lines appreciably broader than H$\beta$, and a strong continuum from young stars.  Many HII regions, and an general outflow is seen in H$\alpha$, \NII\ and \SII. This extends across the field, and looks shock-excited (lines are broad and complex).\newline	
{\bf NGC 7582: } This galaxy displays a star-forming disk \citep{CidFernandes01} out of which a high excitation, bright, and extended ($ > 15$ arcsec, 1.3kpc) ionisation cone with an opening angle of 110$\degr$ emerges . The counter-cone also appears (although heavily reddened) on the far side of the disk, and extends across the field. The ionisation cones appear to be in outflow -see also \citet{Morris85}. Many fine \HII\ regions are embedded in the star forming disk.  An HST red continuum image is available \citep{Malkan98}, , and long slit spectroscopy has been obtained by \citet{Fraquelli03}. \newline		
{\bf NGC 7591: } The nucleus is a heavily reddened LINER with strong old star continuum embedded in a LIRG with many \HII\ regions.  \citet{Zaw09} used data from \citet{Moustakas06} to argue that it is simply a star forming LIRG. However, this appears not to be the case. \newline
{\bf NGC 7590:	} A faint and compact Seyfert 2 nucleus with no apparent ENLR. Many bright HII regions are seen in the surrounding star-forming disk.  \newline
{\bf NGC 7679: }  This Seyfert ~2 galaxy with very narrow emission lines shows a very extended one-sided ionisation cone emerging from a bright central disk of \HII\ regions. The \lOIII\ line profiles in the ENLR are complex.\newline
{\bf NGC 7714: }   This starburst nucleus galaxy is in interaction with the nearby galaxy NGC 7715. The nuclear spectrum shows a pure starburst, with a strong blue continuum of young stars. \ion{He}{2} emission is very clear - see also \citet{Gonzalez95}. This galaxy has been imaged in the red continuum with HST \citep{Malkan98}.\newline

\section{Conclusions}\label{conclusions}
In this paper we have presented images and nuclear spectroscopy  for 64 galaxies drawn from the first data release of the \emph{Siding Spring Southern Seyfert Spectroscopic Snapshot Survey} (S7),  {which has been made available at \url{http://miocene.anu.edu.au/S7/Data_Release_1/} This dataset } demonstrates the utility of integral field spectroscopy of AGN with both a reasonably wide field, good spectral resolution, and a full spectral coverage in the optical. From these data we have examined the spatial relationship between the extended narrow line region and the \HII\ regions in the disks of the host galaxies. {Nearly all of our Seyferts are found to have ENLR. We find that the ENLR that the ENLR lies inside a circumnuclear ring of star-formation in $\sim 25$\% of the galaxies of the sample. This star forming ring is typically a few kiloparsecs across, and  we may tentatively identify this feature as an Inner Lindblad Resonance. We will investigate this point on the basis of the detailed dynamics in a future paper. Where the angle of the ring to the line of sight permits, we find that the ENLR ionisation cone is perpendicular to the star-forming ring, suggesting that in general, the equatorial axis of the inner accretion torus defining the ionisation cones must be well-aligned with the equatorial axis of the outer star-forming ring.
A full examination of the dynamical structure of the ENLR, its zone of influence, the chemical abundances in the circum-nuclear gas, and the constraints that can be placed on the luminosity and EUV spectral energy distribution of the central engine are deferred to future publications.}

As far as the nuclear spectroscopy is concerned, we have found interesting correlations within the class of objects which display Coronal line emission from species such as [\ion{Fe}{7}], [\ion{Fe}{10}] and [\ion{Fe}{14}]. We conclude that their properties are consistent with the model advocated by \citet{Mullaney09} in which coronal lines arise in a dense gas launched from the dusty inner torus ($10^{17} < R/{\rm cm} < 10^{18}$) at very high local ionisation parameter $\log U \sim -0.4$, which is accelerated by radiation pressure to a terminal velocity of a few hundred km~s$^{-1}$. 
For the Type 2 Seyferts, the nuclear spectra are all located in the upper region of the Seyfert sequence defined by the AGN drawn from the SDSS galaxies on the \citet{Baldwin81} and \citet{Veilleux87} diagnostic plots. We conclude that the SDSS galaxies populating the lower region of these diagrams are likely to be composite spectra caused by mixing between the Seyfert nucleus and the surrounding \HII\ regions in the galaxy, {and therefore that the position of these objects on these diagnostic plots is likely a consequence of  aperture effects in the SDSS spectra. }

\acknowledgments 
Dopita and Kewley acknowledge the support of the Australian Research Council (ARC) through Discovery project DP130103925, Dopita also acknowledges financial support from King Abdulaziz University. This research was started as  consequence of the grant to Dopita of an Australian Academy of Science Australia-India Visiting Fellowship (2012-2013). Juneau acknowledges support from the European Research Council through grant ERC-StG- 257720. J.S. acknowledges the European Research Council for the Advanced Grant Program Number 267399-Momentum. Scharw\"achter acknowledges support from the European Research Council through grant ERC-StG- 257720. This research has made use of the NASA/IPAC Extragalactic Database (NED), which is operated by the Jet Propulsion Laboratory, California Institute of Technology, under contract with the National Aeronautics and Space Administration, the NASA Astrophysics Data System (ADS), and SAOImage DS9 (Joye \& Mandel 2003), developed by the Smithsonian Astrophysical Observatory. Last, but not least, we thank the anonymous referee for helping us to much improve this paper.

\bibliographystyle{apj}
\bibliography{ms}

%\begin{table*}
%\centering
%\small
%\caption{Log of the observations}. \label{table:Obs}
%\begin{tabular}{lcccllclc}
%\tableline \tableline
%& & & & & & & &\\
%Galaxy: &  RA & Dec. & $z$ & Type \footnote{As determined from nuclear spectra presented here.} & Type  & Obs. & Observation & Integration \\
%&   \multicolumn{2}{c}{(J2000)} & & & (Veron) & PA$\degr$ & Date  & Time (s) \\
%\tableline
%& & & & & & & &\\

\newpage
\begin{deluxetable}{lcccccccllc}
 \centering
\tabletypesize{\scriptsize}
\tablecaption{Log of the observations. \label{table:Obs}}
\tablewidth{525pt}
\tablehead{
\colhead{Object} & \colhead{RA}         & \colhead{Dec}         &  \colhead{$z$} & \colhead{Type} & \colhead{Type} & \colhead{PA$\degr$} & \colhead{ Date of}  & \colhead{Exposure} & \colhead{Seeing}\\
\colhead{Name} & \colhead{ (J2000)} & \colhead{ (J2000)} &  \colhead{}       & \colhead{}          & \colhead{Veron} & \colhead{} & \colhead{ Observation}  & \colhead{Time (s)} & \colhead{( arcsec)}
}
\startdata
PKS 0056-572    & 00 58 46.7 & -56 59 09.9 & 1.46         & QSO  & S1 & 0 & 2013 Nov 1 & 3000 & 1.1 \\
NGC 424	            & 01 11 27.5 & -38 05 00.7 & 0.0118    & Sey 2,Cor!{$^1$},B(H$\alpha$){$^2$} & S1h & 90  & 2013 Aug 12 & 3000 & 1.8 \\
NGC 613	            & 01 34 18.1 & -29 25 03.0 & 0.0049    & Sey 2 & S? & 90  & 2013 Nov 3 & 3000 & 2.2 \\
IC 1657	            & 01 14 07.0 & -32 39 03.0 & 0.0119    & Sey 2 & S2 & 0  & 2013 Aug 8 & 3000 & 2.1 \\
MARK 573           & 01 43 57.8 & 02 21 03.0  & 0.0172    & Sey 2, Cor.{$^3$}, B(H$\alpha$) & S1h & 90  & 2013 Nov 4 & 3000 & 2.0 \\
IRAS 01475-0740 & 01 50 02.6 & -07 25 50.0 & 0.0177 & Sey 2 & S1h & 90 & 2013 Aug 9 & 3000 & 1.8 \\
NGC 833	            & 02 09 20.8 & -10 07 59.3 & 0.0129    & LINER & S2 & 90 & 2013 Aug 4 & 3000 & 3.0 \\
NGC 835	            & 02 09 24.6 & -10 08 08.3 & 0.0129    & LINER & S2 & 0 & 2013 Aug 3 & 3000 & 1.6 \\
IC 1816	            & 02 31 51.1 & -36 40 11.9 & 0.0169    & Sey 2 & S2 & 90 & 2013 Nov 4 & 3000 & 2.7 \\
NGC 1052           & 02 41 04.9 & -08 15 21.0 & 0.0050    & LINER  & S3b & 90 & 2013 Nov 2 & 3000 & 1.5 \\
NGC 1097           & 02 46 19.2 & -30 16 19.0 & 0.0042    & LINER & S3b & 0 & 2013 Nov 4 & 3000 & 2.7 \\
NGC 1125           & 02 51 40.3 & -16 39 04.0 & 0.0109    & Sey 2 & S2 & 90 & 2013 Nov 1 & 3000 & 1.2 \\
NGC 1204           & 03 04 40.1 & -12 20 22.9 & 0.0154    & SB{$^4$} & S2 & 60 & 2013 Nov 3 & 3000 & 2.3 \\
NGC 1566           & 04 20 01.0  & -54 56 16.7  & 0.0050 & Sey 1, B(H$\alpha$) & S1.5 &  90 & 2013 Nov 1 & 3000 & 1.3 \\
ESO 202-G23     & 04 28 00.0  & -47 54 24.0  & 0.0165 & LINER or Sy 2 & S3 & 0 & 2013 Nov 3 & 3000 & 2.5 \\
NGC 1808           & 05 07 42.1  & -37 30 33.8  & 0.0033 & SB & H2 & 0 & 2013 Nov 2 & 3000 & 1.1 \\
MARK 1210         & 08 04 05.9  &  05 06 49.3  & 0.0135 & Sey 2, Cor., B(H$\alpha$) & S2 & 90  & 2014 Mar 2 & 2700 & 1.6 \\
NGC 2617            & 08 35 38.8  & -04 05 18.0  & 0.0142 & Sey 1 & S1.8 & 90 & 2014 Mar 2 & 2700 & 1.6 \\
MCG -01-24-012 & 09 20 46.3  & -08 03 22.0  & 0.0196 & Sey 2 & S2 & 45  & 2014 Apr 5 & 3000 & 1.7 \\
MCG -05-23-004 & 09 31 07.4  & -30 21 30.3  & 0.0086 & LINER & S3 & 90 & 2014 Apr 5 & 3000 & 1.2 \\
MCG -05-23-008 & 09 44 13.4  & -28 50 55.0  & 0.0084 & LINER & S & 90 & 2014 Apr 8 & 1800 & 1.7 \\
NGC 2992            & 09 45 42.0  & -14 19 33.4  & 0.0077 & Sey 2, B(H$\alpha$) & S1.9 & 45 & 2014 Apr 5 & 3000 & 2.0 \\
MARK 1239         & 09 52 19.1  & -01 36 42.8  & 0.0199 & Sey 1,Cor!, B(H$\alpha$) & S1n & 0 & 2014 Mar 2 & 2700 & 1.8 \\
NGC 3100            & 10 00 40.9  & -31 39 52.0  & 0.0088 & LINER & S3 & 0 & 2014 Apr 5 & 2700 & 2.5 \\
IC 2560                 & 10 16 18.7  & -33 33 50.0  & 0.0097 & Sey 2, Cor., B(H$\alpha$) & S2 & 45  & 2014 Apr 8 & 1800 & 1.5 \\
MCG -06-23-038 & 10 29 45.6  & -38 20 55.0  & 0.0152 & Sey 2 & S & 90 & 2014 Mar 2 & 2700 & 2.3 \\
IRAS 11215-2806 & 11 24 09.8  & -28 23 49.0  & 0.0140& Sey 2 & S2 & 135 & 2014 Apr 7 & 2700 & 2.0 \\
NGC 3783            & 11 39 02.2  & -37 44 17.0   & 0.0090 & Sey 1, Cor! & S1.5 & 0 & 2014 Apr 7 & 1800 & 2.5 \\
NGC 4303            & 12 21 55.4  & 04 28 31.0   & 0.0052 & SB or LINER & S2 & 0 & 2014 & 1800 &  1.3 \\
NGC 4404            & 12 26 16.2  & -07 40 51.0  & 0.0186 & LINER & S3 & 90 & 2014 Apr 5 & 3000 & 1.8 \\
3C 278                  & 12 54 36.9  & -12 33 28.6  & 0.0150 & Elliptical & S3 & 90 & 2014 Apr 5 & 3000 &   2.1 \\
NGC 5253            & 13 39 56.0  & -31 38 13.0  & 0.0014 & SB & H2  & 0 & 2013 Aug 5 & 3000{$^5$} & 1.9 \\
NGC 5506            & 14 13 14.8  & -03 12 27.0  & 0.0062 & Sey 2, Cor & S1i & 90 & 2013 Aug 6 & 3000 & 0.9 \\
NGC 5597            & 14 24 28.0  & -16 45 44.0  & 0.0089 & SB + WR{$^6$} & H2 & 45 & 2014 Apr 8 & 1800 & 1.1 \\
NGC 5664            & 14 33 43.7  & -14 37 10.5  & 0.0152 & SB + Sey 2 & S2 & 30 & 2014 Apr 5 & 3000 & 2.0 \\
NGC 5728            & 14 42 24.7  & -17 15 07.0  & 0.0093 & Sey 2,Cor & S1.9 & 120 & 2014 Apr 8 & 1800 & 1.2 \\
NGC 5757            & 14 46 46.4  & -19 04 42.7  & 0.0089 & SB & H2 & 0 & 2013 Aug 3 & 3000 & 2.0 \\
NGC 6000            & 15 49 49.6  & -29 23 13.0  & 0.0073 & SB & H2 & 0 & 2014 Apr 5 & 3000 & 1.8 \\
ESO 137-G34      & 16 35 13.9  & -58 04 48.3  & 0.0091 & Sey 2 & S2 & 90 & 2013 Aug 10 & 3000 & 1.4 \\
ESO 138-G01      & 16 51 20.1  & -59 13 48.0  & 0.0091 & Sey 2, Cor.,B(H$\alpha$) & S2 & 90 & 2014 Apr 8 & 1800 & 1.1 \\
NGC 6221            & 16 52 46.0  & -59 13 01.0  & 0.0050 & SB + Sey 2 & S2 & 90 & 2014 Apr 8 & 1800 & 1.2 \\
NGC 6300            & 17 16 59.3  & -62 49 14.2  & 0.0037 & Sey 2 & S2 &90  & 2013 Aug 10 & 3000 & 1.4 \\
FAIRALL 49         & 18 36 58.2 & -59 24 08.0   & 0.0200 & Sey 2, B(H$\alpha$) & S1h & 90 & 2014 Apr 8 & 1800 & 1.2 \\
ESO 103-G35      & 18 38 20.4  & -65 25 38.4  & 0.0130 & Sey 2 &S2  & 90 & 2013 Aug 9 & 3000 & 1.5 \\
\hline\\
FAIRALL 51         & 18 44 53.8  & -62 21 51.0  & 0.0142 & Sey 1, Cor! & S1.5 & 0  & 2013 Aug 6 & 3000 & 1.3 \\
NGC 6812            & 19 45 24.8  & -55 20 50.2  & 0.0154 & LINER & S? & 90 & 2013 Aug 4 & 3000 & 1.6 \\
ESO 339-G11      & 19 57 37.5  & -37 56 08.0  & 0.0192 & Sey 2, Cor., B(H$\alpha$) & S2 & 90& 2013 Aug 5 & 3000 & 1.5 \\
NGC 6860            & 20 08 47.0  & -61 06 01.0  & 0.0149 & Sey 1 &  S1.5 & 0  & 2013 Aug 12 & 3000 & 2.6 \\
NGC 6890            & 20 18 18.1  & -44 48 24.2  & 0.0081 & Sey 2, Cor. & S1.9 & 0 & 2013 Aug 10 & 3000 & 1.1 \\
NGC 6915            & 20 27 46.1  & -03 04 37.6 & 0.0189 & LINER & S3 & 90 & 2013 Aug 5 & 3000 & 1.5 \\
NGC 6926            & 20 33 06.2  & -02 01 37.8 & 0.0196 & Sey 2 & S2 & 0  & 2013 Aug 9 & 3000 & 1.1 \\
IC5063                  & 20 52 02.2  & -57 04 07.5  & 0.0113 & Sey 2, B(H$\alpha$) & S1h & 90 & 2013 Aug 12 & 3000 & 2.0 \\
IC 1368                  & 21 14 12.6  & 02 10 40.7   & 0.0130 & Sey 2 & S2 &  90 & 2013 Aug 6 & 3000 & 1.5 \\
NGC 7130/IC 5135   & 21 48 19.4  & -34 57 03.3  & 0.0161 & SB + Sey 2 & S1.9 & 90  & 2013 Aug 3 & 3000 & 1.3 \\
NGC 7213             & 22 09 16.2  & -47 10 00.7  & 0.0058 & LINER & S3b & 90 & 2013 Aug 5 & 3000 & 1.3 \\
IC 1459                  & 22 57 10.5  & -36 27 45.0  & 0.0060 & LINER & S3 & 0 & 2013 Aug 3 & 3000 & 1.7 \\
NGC 7469             & 23 03 16.0  & 08 52 24.5   & 0.0163 & Sey 1, Cor. & S1.5 & 90 & 2013 Aug 10 & 3000 & 1.3 \\
NGC 7496             & 23 09 47.3  & -43 25 40.5  & 0.0055 & SB + Sey 2 &  S2 &  0 & 2013 Nov 1 & 3000 & 1.2 \\
NGC 7552             & 23 16 10.8  & -42 35 05.0  & 0.0054 & SB & H2 & 90 & 2013 Aug 6 & 3000 & 1.5 \\
NGC 7591             & 23 18 16.3  & 06 35 10.0   & 0.0165 & LINER & S & 0 & 2013 Aug 5 & 3000 & 1.1 \\
NGC 7582             & 23 18 23.4  & -42 22 13.6 & 0.0053 & Sey 2 + SB & S1i & 0 & 2013 Nov 2 & 3000 & 1.3 \\
NGC 7590             & 23 18 54.5  & -42 14 07.8 & 0.0052 & Sey 2 & S2 &  0 & 2013 Nov 3 & 3000 &  1.5 \\
NGC 7679             & 23 28 46.8  & 03 30 45.0  & 0.0171 & SB + Sey 2 & S1.9 & 90 & 2013 Nov 4 & 3000 & 1.8 \\
NGC 7714             & 23 36 14.2  & 02 09 17.6  & 0.0093 & SB & H2 & 0 & 2013 Aug 5 & 3000 & 1.5 \\
\enddata
\newline
$^1$Strong and high-excitation coronal lines present.
$^2$Broad component detected in H$\alpha$
$^3$Coronal lines present.\\
$^4$SB = Starburst.
$^5$Bright lines saturated.
$^6$WR = Wolf Rayet features present.
\end{deluxetable}

\newpage
\begin{deluxetable}{lccccccccr}
\rotate
\small
 \centering
\tabletypesize{\scriptsize}
\tablecaption{Derived Nuclear Extinctions, H$\beta$ and \OIII\ Fluxes, Luminosities and extent of the narrow-line region \label{table:Flux}}
\tablewidth{535pt}
\tablehead{
\colhead{Object}  & \colhead{Type} & \colhead{ D$_{\rm Lum.}$}  &  \colhead{$A_{\rm V}$}  &  \colhead{Aperture}  &  \colhead{$F_{\rm H\beta}$}  &  \colhead{$F_{\rm O~III}$}  &  \colhead{$\log L_{\rm H\beta}$}  &  \colhead{$\log L_{\rm O~III}$}  &  \colhead{Size ENLR}  \\
\colhead{}  & \colhead{} & \colhead{(Mpc)}  &  \colhead{(mag.)}  &  \colhead{(kpc)}  &  \colhead{(erg~cm$^{-2}$s$^{-1}$)}  &  \colhead{} &  \colhead{(erg~s$^{-1}$)}  &  \colhead{}  &  \colhead{( arcsec)}  \\
}
\startdata

%\begin{table*}
%\small
%\centering
%\caption{Derived Nuclear Extinctions, H$\beta$ and \OIII\ Fluxes, Luminosities and extent of the narrow-line region.} \label{table:Flux}
%\begin{tabular}{lccccccccr}
%\tableline \tableline
%&  & & & & & & & & \\
%Object:  & Type:  & D$_{\rm Lum.}$\footnote{Luminosity Distance from NED} &  $A_{\rm V}$ & Aperture\footnote{Diameter of aperture from which the nuclear spectrum \newline was extracted (4 arcsec)} & $F_{\rm H\beta}$ & $F_{\rm O~III}$ & $\log L_{\rm H\beta}$ & $\log L_{\rm O~III}$ & Size ENLR\footnote{Approximate detectable extent of \OIII\ emission region.}\\
% &  & (Mpc) & (mag.) & (kpc) & \multicolumn{2}{c}{(erg~cm$^{-2}$s$^{-1}$)}  & \multicolumn{2}{c}{(erg~s$^{-1}$)} & ( arcsec)\\
%\tableline
%&  & & & & & & & & \\
NGC 424                & Seyfert 2 & 45.6 & 1.76 & 0.87 & 5.12E-14 & 3.19E-13 & 40.87 & 41.67 & $18\times14$\\
NGC 613                & Seyfert 2 & 17.1 & 2.73 & 0.32 & 3.44E-14 &  1.78E-14 & 40.27  & 39.99 & $20\times14$\\
IC 1657                   & Seyfert 2 & 46.2 & 1.65 & 0.88 & 3.56E-15 & 6.40E-15 & 39.68 & 39.93 & $11\times10$\\
MARK 573            & Seyfert 2 & 67.4 & 0.74 & 1.28 & 4.49E-14 & 4.60E-13 & 40.71 & 41.72 & $>25\times15$\\
IRAS 01475-074  & Seyfert 2 & 69.7 & 2.93 & 1.32 & 9.72E-15 & 4.86E-14 & 41.03 & 41.73 & $10\times7$\\
NGC 833               & LINER  & 50.0 & 0.00 & 0.95 & 5.95E-15 & 7.26E-15 & 39.25 & 39.33 & \\
NGC 835               & Seyfert 2  & 52.9 & 1.82 & 1.01 & 1.22E-14 & 5.56E-15 & 40.40 & 40.06 & \\
IC 1816                 & Seyfert 2 & 68.4 & 0.72 & 1.30 & 1.56E-14 & 1.56E-13 & 40.25 & 41.25 & $15\times13$\\
NGC 1052            & LINER  & 17.8 & 0.04 & 0.34 & 5.95E-14 & 1.07E-14 & 39.37 & 38.62 & \\
NGC 1097            & LINER  & 15.2 & 0.45 & 0.29 & 1.05E-14 & 8.17E-15 & 38.66 & 38.55 & \\
NGC 1125            & Seyfert 2  & 42.6 & 2.46 & 0.81 & 2.00E-14 & 1.23E-13 & 40.71 & 41.50 & $\sim30\times10$\\
NGC 1204            & SB & 61.4 & 4.64 & 1.17 & 1.99E-15 & 7.38E-16 & 40.98 & 40.55 & \\
NGC 1566             & Seyfert 1 & 20.5 & 0.09 & 0.39 & 6.47E-14 & 1.89E-13 & 39.55 & 40.01 & $15\times10$\\
ESO 202-G23       & Seyfert 2  & 68.5 & 2.56 & 1.30 & 1.52E-15 & 2.74E-15 & 40.05 & 40.30 & \\
NGC 1808             & SB  & 14.0 & 3.38 & 0.27 & 4.34E-14 & 1.11E-14 & 40.48 & 39.89 & \\
MARK 1210          & Seyfert 2 & 59.5 & 1.03 & 1.13 & 5.98E-14 & 5.74E-13 & 40.85 & 41.83 & $<5\times5$\\
NGC 2617             & Seyfert 1  & 61.2 & 2.55 & 1.16 & 1.84E-14 & 2.89E-14 & 41.03 & 41.23 & Knot at 15\\
MCG -01-24-012  & Seyfert 2 & 86.6 &  & 1.65 &  &  &  &  & $9\times6$\\
MCG -05-23-004  & LINER  & 39.9 & 0.00 & 0.76 & 4.38E-15 & 5.45E-15 & 38.92 & 39.01 & \\
MCG -05-23-008  & LINER  & 39.3 & 0.00 & 0.75 & 4.56E-15 & 3.70E-15 & 38.92 & 38.83 & \\
NGC 2992             & Seyfert 2 & 36.6 & 3.78 & 0.70 & 1.55E-14 & 8.15E-14 & 41.05 & 41.77 & $>35\times25$\\
MARK 1239          & Seyfert  1 & 88.0 & 2.58 & 1.67 & 1.39E-13 & 2.39E-13 & 42.23 & 42.47 & $17\times15$\\
NGC 3100             & LINER  & 40.9 & 0.00 & 0.78 & 9.43E-15 & 8.47E-15 & 39.27 & 39.23 & \\
IC 2560                  & Seyfert 2 & 44.9 & 1.33 & 0.85 & 2.30E-14 & 2.46E-13 & 40.32 & 41.35 & $13\times18$\\
MCG -06-23-038  & Seyfert 2 & 67.3 & 2.47 & 1.28 & 5.50E-15 & 3.93E-14 & 40.55 & 41.41 & $11\times9$\\
IRAS 11215-2806 & Seyfert 2 & 62.9 & 1.58 & 1.20 & 4.05E-15 & 3.19E-14 & 39.97 & 40.87 & $\sim7\times7$\\
NGC 3783             & Seyfert 1 & 44.7 & 0.36 & 0.85 & 4.03E-14 & 3.82E-13 & 40.14 & 41.12 & $\sim18\times18$\\
NGC 4303             & SB or LINER & 26.3 & 1.33 & 0.50 & 3.59E-14 & 3.31E-14 & 40.05 & 40.02 & \\
NGC 4404             & LINER  & 82.6 & 0.00 & 1.57 & 4.65E-15 & 3.25E-15 & 39.58 & 39.42 & \\
NGC 5506             & Seyfert 2  & 29.1 & 2.07 & 0.55 & 4.85E-14 & 3.37E-13 & 40.60 & 41.44 & $15\times12$\\
NGC 5597             & SB + WR & 40.5 & 1.44 & 0.77 & 7.75E-14 & 2.48E-14 & 40.81 & 40.31 & \\
NGC 5664             & SB + Seyfert 2 & 66.3 & 1.94 & 1.26 & 7.68E-15 & 2.27E-14 & 40.45 & 40.92 & $>25\times25$ \\
\hline\\
NGC 5728             & Seyfert 2  & 41.9 & 1.23 & 0.80 & 3.48E-14 & 3.31E-13 & 40.40 & 41.38 & Jets, $22\times7$\\
NGC 5757             & SB  & 40.0 & 2.14 & 0.76 & 3.38E-14 & 4.97E-15 & 40.75 & 39.91 & \\
NGC 6000             & SB  & 32.1 & 3.14 & 0.61 & 3.70E-14 & 6.62E-15 & 41.03 & 40.28 & \\
ESO 137-G34       & Seyfert 2  & 38.8 & 2.16 & 0.74 & 3.12E-14 & 2.90E-13 & 40.69 & 41.66 & $>40\times25$\\
ESO 138-G01       & Seyfert 2 & 38.6 & 2.12 & 0.73 & 6.98E-14 & 5.75E-13 & 41.02 & 41.93 & Jets$ > 35$\\
NGC 6221             & SB + Sey & 21.4 & 3.49 & 0.41 & 8.29E-14 & 4.25E-14 & 41.18 & 40.89 & $<5\times5$\\
NGC 6300             & Seyfert 2  & 15.7 & 4.33 & 0.30 & 2.90E-15 & 4.29E-14 & 39.82 & 40.99 & $\sim7\times5$\\
FAIRALL 49          & Seyfert 2 & 83.1 & 3.31 & 1.58 & 5.53E-14 & 2.09E-13 & 42.11 & 42.68 & $28\times10$\\
ESO 103-G35       & Seyfert 2  & 77.7 & 2.36 & 1.48 & 6.18E-15 & 4.55E-14 & 40.68 & 41.54 & $\sim26 \times8$\\
FAIRALL 51          & Seyfert 1 & 58.5 &  0.98 & 1.11 &  3.30e-14 &  6.09 e-14 &  40.56 & 40.83  & $10\times7$\\
NGC 6812             & LINER & 62.9 & 0.05 & 1.20 & 6.00E-15 & 9.11E-15 & 39.47 & 39.66 & \\
ESO 339-G11       & Seyfert 2 & 77.7 & 3.86 & 1.48 & 6.35E-15 & 4.99E-14 & 41.35 & 42.24 & $\sim9\times5$\\
NGC 6860             & Seyfert 1  & 60.7 & 1.04 & 1.15 & 1.02E-14 & 5.38E-14 & 40.11 & 40.83 & $15\times15$\\
NGC 6890             & Seyfert 2 & 31.2 & 1.50 & 0.59 & 1.30E-14 & 1.42E-13 & 39.83 & 40.87 & $\sim8\times8$\\
NGC 6915             & LINER & 74.9 & 0.00 & 1.42 & 2.88E-15 & 3.25E-15 & 39.28 & 39.34 & $7times5$ \\
NGC 6926             & Seyfert 2  & 82.6 & 3.92 & 1.57 & 8.98E-16 & 3.94E-15 & 40.58 & 41.22 & $<3\times3$ \\
IC5063                   & Seyfert 2 & 45.3 & 2.28 & 0.86 & 3.07E-14 & 2.41E-13 & 40.87 & 41.77 & $>35\times25$\\
IC 1368                  & Seyfert 2 & 49.6 & 5.07 & 0.94 & 1.79E-15 & 6.46E-15 & 40.94 & 41.49 & $<5\times5$\\
NGC 7130             & SB + Seyfert & 63.6 & 1.79 & 1.21 & 3.35E-14 & 1.55E-13 & 40.99 & 41.66 & $9\times9$\\
NGC 7213             & LINER & 21.2 & 0.17 & 0.40 & 6.81E-14 & 8.17E-14 & 39.64 & 39.72 &  $9\times9$ \\
IC 1459                  & LINER & 21.1 & 0.00 & 0.40 & 3.66E-14 & 2.34E-14 & 39.29 & 39.09 & \\
NGC 7469             & Seyfert 1 & 62.7 & 3.47 & 1.19 & 7.44E-14 & 3.88E-13 & 42.06 & 42.77 & $16\times8$\\
NGC 7496             & SB + Sey & 19.4 & 1.45 & 0.37 & 7.07E-14 & 3.89E-14 & 40.14 & 39.88 & $<6\times6$\\
NGC 7591             & LINER & 63.6 & 3.85 & 1.21 & 1.38E-15 & 9.98E-16 & 40.51 & 40.37 & \\
NGC 7582             & Seyfert 2 + SB  & 18.3 & 3.11 & 0.35 & 5.33E-14 & 2.93E-14 & 40.69 & 40.43 & \\
NGC 7590             & Seyfert 2  & 18.3 & 1.89 & 0.35 & 6.03E-15 & 2.09E-14 & 39.21 & 39.75 & $<3\times3$\\
NGC 7679             & SB + Seyfert & 66.2 & 1.03 & 1.26 & 6.13E-14 & 6.44E-14 & 40.95 & 40.98 & $\sim20\times5$\\
NGC 7714             & SB  & 33.5 & 1.49 & 0.64 & 2.88E-13 & 4.01E-13 & 41.24 & 41.38 & \\
%\tableline
%\end{tabular}
%\end{table*}
\enddata
\end{deluxetable}

\newpage
\begin{deluxetable}{lccclccc}
\small
\centering
\tabletypesize{\scriptsize}
\tablecaption{ Velocity widths (km~s$^{-1}$) of the three fitted components to the emission lines. Note that these are not corrected for the instrumental width ($\sim 45$km~s$^{-1}$). \label{Velocity}}
\tablewidth{300pt}
\tablehead{
\colhead{Object } & \colhead{$V_1$} &  \colhead {$V_2$} & \colhead{$V_3$} & \colhead{Object } & \colhead{$V_1$}  & \colhead {$V_2$} &  \colhead{$V_3$} \\
}

\startdata
NGC 424 & 182 & 385 & -- & NGC 5664 & 113 & --  &  --\\
NGC 613 & 70 & 198 & -- & NGC 5728 & 113 & -- & --\\
IC 1657 & 38 & 124 & -- & NGC 5757 & 64 & 153 & --\\
MARK 573 & 97 & 237 & -- & NGC 6000 & 120 & 241 &-- \\
IRAS 01475 & 62 & 207 &  --& ESO 127-G34 & 131 & 137 & 392\\
NGC 833 & 236 & -- & -- & ESO 138-G01 & 88 & 211 & --\\
IC1816 & 122 & 127 & -- & NGC 6221 & 51 & 210 & --\\
NGC 1052 & 158 & 429 & -- & NGC 6300 & 93 & 220 & --\\
NGC 1097 & 184 & -- & -- & FAIRALL 49 & 77 & 150 & 494\\
NGC 1125   & 114 & 344 & -- & ESO 103-G35 & 124 & 167 & 459\\
NGC 1204 & 41 & 123 &  -- & FAIRALL 51 & 106 & 705 & 811\\
NGC 1566  & 83 & 140 & 684 & NGC 6812 & 269 & -- & --\\
ESO 202-G23  & 241 & -- &  -- & ESO 339-G11 & 107 & 284 & -- \\
NGC 1808 & 78 & 180 & -- & NGC 6860 & 104 & 229 & 1447\\
MARK 1210  & 121 & 475 & -- & NGC 6890 & 61 & 148 & 354\\
NGC 2617 & 120 & 1365 & -- & NGC 6915 & 168 & -- & -- \\
MCG-05-23-004 & 158 & -- & -- & NGC 6926 & 228 & -- & -- \\
MCG-05-23-008 & 211 & -- & -- & IC 5063 & 154 & 363 & \\
NGC 2992 & 73 & 204 & -- & IC1368 & 72 & 98 &  --\\
MARK 1239 & 148 & 534 & -- & NGC 7130 & 70 & 82 & 421\\
NGC 3100 & 120 & 589 & -- & NGC 7213 & 131 & 625 & -- \\
IC 2560 & 81 & 212 & -- & IC 1459 & 173 & 631 & -- \\
MCG -06-23-038 & 133 & 312 & -- & NGC 7469 & 127 & 229 & 958\\
IRAS 11215 & 66 & 300 & -- & NGC 7496 & 50 & 72 & 234\\
NGC 3783 & 78 & 213 & 1195 & NGC 7591 & 194 &  -- & -- \\
NGC4303 & 52 & 231 & -- & NGC 7582 & 64 & 138 & -- \\
NGC4404 & 309 & -- & -- & NGC 7590 & 65 & -- & -- \\
3C278 & 421 & -- & -- & NGC 7679 & 89 & 397 & -- \\
NGC 5506 & 95 & 190 & 426 & NGC 7714 & 66 & 113 &-- \\
NGC 5597 & 41 & 109 & -- &  &  &  & \\
\enddata
\end{deluxetable}

\newpage
\begin{deluxetable}{llrrrrrrrr}
\small
\rotate
 \centering
\tabletypesize{\scriptsize}
\tablecaption{Measured de-reddened line fluxes relative to H$\beta$ = 100\label{Lines}}
\tablewidth{590pt}
\tablehead{
%\colhead{Lambda} \colhead{Line ID}  \colhead {A} \colhead{B} \colhead{C} \colhead{D} \colhead{E} \colhead{F} \colhead{G} \colhead{H} \\
}
\startdata
Lambda  & Line  ID & NGC 424 & NGC 613 & IC 1657 & MARK 573 & IRAS 01475 & NGC 833 & IC 1816 & NGC 1052 \\
\hline\\
3586.3  & [Fe VII]  & 20.1   $\pm$ 1.4 & --   &  --  & 4.9     $\pm$ 1.2  & 6.5      $\pm$ 5.6 & --    & --     & -- \\   
3726,9 & [O II]       & 113.8$\pm$ 3.1  & 242.8 $\pm$ 8.1  & 367.7 $\pm9.6$ & 213.5 $\pm5.4$ & 178.2  $\pm$ 6.2 & 223.8  $\pm$ 4.9  & 199.7  $\pm$ 1.0 & 289.1 $\pm$ 4.9 \\
3868.8 & [Ne III] & 92.8     $\pm$ 1.7  & 15.3   $\pm$ 0.9  & --   & 85.8    $\pm9.4$ & 43.0    $\pm$ 12.6 & --   & 100.8  $\pm$ 1.3 & 8.5  $\pm$ 29.9 \\
3889  & H,He I & 27.1     $\pm$ 2.2   & 20.0   $\pm$ 3.7  & --    & 21.2    $\pm1.9$ & 17.8   $\pm$ 1.8 & 45.2    $\pm$ 1.4   & 22.6  $\pm$ 0.9    & 23.2  $\pm$ 29.9\\
3967,70 & [Ne III],H & 44.6  $\pm$ 2.4 & 16.8   $\pm$ 4.0 & -- & 44.5    $\pm$ 1.0 & 28.4  $\pm$ 3.0 & 44.6  $\pm$ 2.1 & 47.2  $\pm$ 0.3 & 28.4  $\pm$ 14.3\\
4026.2 & He I     & -- & -- & -- & 2.7     $\pm$ 0.7 & -- & -- & -- & --\\
4068,76 & [S II]  & 16.5  $\pm$ 0.7 & 8.0    $\pm$ 1.2 & -- & 16.3    $\pm$ 1.9 & 18.9  $\pm$ 3.4 & 16.0  $\pm$ 5.9 & 53.4  $\pm$ 1.0 & 37.0  $\pm$ 1.3\\
4101.7 & H$\delta$ & 31.2   $\pm$ 1.5 & 29.4   $\pm$ 9.0 & -- & 28.5    $\pm$ 1.4 & 34.7  $\pm$ 4.6 & 56.6  $\pm$ 1.9 & 26.7  $\pm$ 0.6 & 40.8  $\pm$ 1.9\\
4340.5 & H$\gamma$   & 50.7   $\pm$ 2.6 & 51.7   $\pm$ 4.8 & -- & 51.2    $\pm$ 2.4 & 51.1  $\pm$ 6.4 & 80.9  $\pm$ 3.6 & 48.3  $\pm$ 0.6 & 70.6  $\pm$ 2.7\\
4363.2 & [O III]  & 37.3   $\pm$ 1.5 & 0.2    $\pm$ 0.1 & -- & 14.5    $\pm$ 1.1 & 22.3  $\pm$ 1.8 & 0.4  $\pm$ 0.2 & 21.7  $\pm$ 0.5 & 14.9  $\pm$ 2.2\\
4471.5 & He I    & 4.1    $\pm$ 0.3 & 2.4    $\pm$ 2.0 & -- & 3.8     $\pm$ 0.3 & 7.7  $\pm$ 3.1 & 12.7  $\pm$ 5.0 & 6.9  $\pm$ 1.4 & 4.9  $\pm$ 1.0\\
4685.7 & He II   & 32.4   $\pm$ 0.9 & -- & -- & 29.4    $\pm$ 2.5 & 8.2  $\pm$ 0.8 & -- & 9.6  $\pm$ 0.5 & --\\
4861.3 & H$\beta$   & 100.0  $\pm$ 3.2 & 100.0  $\pm$ 3.6 & 100.0  $\pm$ 4.3 & 100.0   $\pm$ 0.8 & 100.0  $\pm$ 6.5 & 100.0  $\pm$ 1.7 & 100.0  $\pm$ 1.6 & 100.0  $\pm$ 1.6\\
4958.9 & [O III]  & 208.0  $\pm$ 5.5 & 17.3   $\pm$ 0.4 & 60.0   $\pm$ 1.7 & 342.0   $\pm$ 1.7 & 166.6  $\pm$ 5.4 & 40.6  $\pm$ 12.4 & 333.6  $\pm$ 2.8 & 59.9  $\pm$ 2.1\\
5006.8 & [O III]  & 623.9  $\pm$ 16.4& 51.8   $\pm$ 1.3 & 179.9  $\pm$ 5.2 & 1026.0  $\pm$ 5.2 & 499.7  $\pm$ 16.2 & 121.9  $\pm$ 37.3 & 1000.9  $\pm$ 8.4 & 179.7  $\pm$ 6.2\\
5198,200 & [N I] & 3.9  $\pm$ 16.7& -- & -- & 10.7    $\pm$ 2.0 & -- & 28.9  $\pm$ 10.0 & 18.4  $\pm$ 0.6 & 18.8  $\pm$ 1.4\\
5302.6 & [Fe XIV] & 8.1    $\pm$ 0.7 & -- & -- & -- & 1.8  $\pm$ 0.2 & -- & 3.3  $\pm$ 1.2 & --\\
5411.4 & He I     & 2.2    $\pm$ 0.7 & -- & -- & 2.4    $\pm$ 1.0 & -- & -- & -- & --\\
5577.3 & [O I]    & -- & -- & -- & -- & -- & -- & -- & 2.2  $\pm$ 1.3\\
5720.7 & [Fe VII]  & 15.0   $\pm$ 0.8 & -- & -- & 10.8  $\pm$ 2.9 & -- & -- & 13.6  $\pm$ 1.1 &  --\\
5754.9 & [N II]   & -- & -- & -- & 4.9  $\pm$ 1.4 & -- & -- & 19.1  $\pm$ 0.6 & 2.8  $\pm$ 0.5\\
5875.6 & He I   & 10.1   $\pm$ 0.6 & 4.3    $\pm$ 0.3 & -- & 9.1  $\pm$ 1.2 & 12.1  $\pm$ 1.9 & -- & 14.5  $\pm$ 0.6 & --\\
6087.0 & [Fe VII]  & 26.4     $\pm$ 1.1 & -- & -- & 13.9  $\pm$ 1.6 & 3.6  $\pm$ 2.0 & -- & 13.4  $\pm$ 0.8 & --\\
6300.3 & [O I]    & 20.7   $\pm$ 8.6 & 14.1   $\pm$ 0.7 & -- & 30.2  $\pm$ 0.7 & 39.0  $\pm$ 4.9 & 86.9  $\pm$ 4.7 & 83.0  $\pm$ 1.3 & 154.8  $\pm$ 6.9\\
6363.8 & [O I]    & 7.0    $\pm$ 0.6 & 5.6    $\pm$ 0.3 & -- & 8.3  $\pm$ 0.6 & 12.3  $\pm$ 1.5 & 21.4  $\pm$ 9.6 & 17.5  $\pm$ 0.7 & 41.9  $\pm$ 12.7\\
6374.5 &[Fe X]  & 12.1   $\pm$ 0.5 & -- & -- & 5.1  $\pm$ 1.2 & -- & -- & 9.6  $\pm$ 0.6 & 5.8  $\pm$ 0.7\\
6547.9 & [N II]   & 67.5   $\pm$ 3.1 & 56.9   $\pm$ 1.9 & 63.2   $\pm$ 1.9 & 78.2  $\pm$ 1.4 & 56.5  $\pm$ 1.7 & 115.1  $\pm$ 35.2 & 151.3  $\pm$ 1.8 & 120.5  $\pm$ 2.1\\
6562.8 & H$\alpha$   & 286.0  $\pm$ 9.0 & 286.0  $\pm$ 6.5 & 286.0  $\pm$ 9.9 & 286.0  $\pm$ 5.2 & 286.0  $\pm$ 7.5 & 286.0  $\pm$ 15.1 & 286.0  $\pm$ 6.8 & 286.0  $\pm$ 9.9\\
6583.2 &[N II]   & 202.6  $\pm$ 9.3 & 170.8  $\pm$ 5.6 & 189.6  $\pm$ 5.6 & 234.5  $\pm$ 4.1 & 169.5  $\pm$ 5.1 & 345.4  $\pm$ 15.6 & 453.8  $\pm$ 5.4 & 361.5  $\pm$ 6.2\\
6678.2 &He I    & -- & -- & -- & 2.9  $\pm$ 0.7 & 2.6  $\pm$ 6.1 & -- & -- & --\\
6716.4 &[S II]   & 26.1   $\pm$ 2.6 & 51.9   $\pm$ 1.2 & 107.6  $\pm$ 5.0 & 79.5  $\pm$ 1.4 & 30.5  $\pm$ 9.3 & 159.1  $\pm$ 5.1 & 70.5  $\pm$ 2.9 & 153.9  $\pm$ 40.9\\
6730.8 & S II]   & 26.8   $\pm$ 2.6 & 39.4   $\pm$ 1.4 & 77.8   $\pm$ 2.6 & 73.4  $\pm$ 1.4 & 23.5  $\pm$ 7.2 & 111.1  $\pm$ 29.6 & 145.8  $\pm$ 2.5 & 192.9  $\pm$ 31.2\\
\hline\\
Lambda  & Line  ID & NGC 1097 & NGC 1125 & NGC 1204 & NGC 1566 & ESO 202-G23 & NGC 1808 &MARK 1210 & NGC 2617\\
\hline\\
3586.3 & [Fe VII] & -- & -- & -- & 2.4$\pm$ 5.7 & -- & -- & 4.4$\pm$ 0.7 & -- \\ 
3726,9  & [O II] & 136.6$\pm$ 3.2 & 314.1$\pm$ 2.9 & 331.0$\pm$ 22.0 & 95.5$\pm$ 1.4 & 375.2$\pm$ 15.2 & 78.1$\pm$ 1.3 & 89.8$\pm$ 31.6 & 215.3$\pm$ 27.7 \\ 
3868.8 & [Ne III] & -- & 68.8$\pm$ 1.9 & -- & 14.9$\pm$ 1.1 & -- & -- & 116.1$\pm$ 3.4 & 33.2$\pm$ 4.7\\ 
3889  & H,He I & 35.2$\pm$ 3.0 & 21.5$\pm$ 1.0 & 27.1$\pm$ 8.2 & 11.6$\pm$ 0.6 & 18.8$\pm$ 10.8 & 10.7$\pm$ 1.0 & 20.3$\pm$ 2.8 & -- \\ 
3967,70 & [Ne III],H & 29.0$\pm$ 2.5 & 31.1$\pm$ 12.4 & -- & 21.6$\pm$ 0.4 & -- & -- & 53.5$\pm$ 1.1 & -- \\ 
4026.2 & He I & -- & 8.2$\pm$ 5.7 & -- & -- & -- & -- & 1.0$\pm$ 0.2 & -- \\
4068,76 & [S II] & 10.3$\pm$ 8.0 & 19.7$\pm$ 1.7 & 16.8$\pm$ 9.0 & 5.7$\pm$ 0.6 & -- & 7.1$\pm$ 1.7 & 53.6$\pm$ 1.7 & --\\
4101.7 & H$\delta$ & -- & 32.0$\pm$ 5.4 & 32.5$\pm$ 5.7 & 26.3$\pm$ 0.6 & 49.0$\pm$ 5.6 & 20.3$\pm$ 1.3 & 25.1$\pm$ 0.7 & -- \\
4340.5 & H$\gamma$ & -- & 55.8$\pm$ 2.8 & 57.9$\pm$ 4.2 & 48.2$\pm$ 0.4 & 47.5$\pm$ 9.1 & 37.3$\pm$ 4.9 & 46.8$\pm$ 1.3 & 50.9$\pm$ 7.2 \\
4363.2 & [O III] & 5.2$\pm$ 1.6 & 11.3$\pm$ 3.4 & -- & 15.6$\pm$ 0.3 & -- & -- & 38.9$\pm$ 1.3 & -- \\
4471.5 & He I & 6.5$\pm$ 3.0 & -- & 10.1$\pm$ 1.5 & 1.3$\pm$ 0.1 & 13.8$\pm$ 9.2 & -- & 4.6$\pm$ 1.0 & 3.3$\pm$ 2.2\\
4685.7 & He II & -- & -- & -- & --& -- & -- & 19.8$\pm$ 2.9 & --\\
4861.3 & H$\beta$ & 100.0$\pm$ 3.9 & 100.0$\pm$ 1.8 & 100.0$\pm$ 6.8 & 100.0$\pm$ 1.5 & 100.0$\pm$ 3.1 & 100.0$\pm$ 1.4 & 100.0$\pm$ 1.4 & 100.0$\pm$ 50.2\\
4958.9 & [O III] & 27.2$\pm$ 6.2 & 205.4$\pm$ 5.2 & 12.3$\pm$ 1.1 & 61.3$\pm$ 5.3 & 60.3$\pm$ 49.0 & 8.9$\pm$ 0.2 & 320.1$\pm$ 3.1 & 52.5$\pm$ 15.5\\
5006.8 & [O III] & 81.7$\pm$ 18.7 & 616.1$\pm$ 15.6 & 37.0$\pm$ 3.2 & 183.9$\pm$ 16.0 & 181.0$\pm$ 147.0 & 26.6$\pm$ 0.5 & 960.2$\pm$ 9.3 & 157.4$\pm$ 16.5\\
5198,200 & [N I] & 16.9$\pm$ 11.5 & 8.1$\pm$ 0.4 & 15.9$\pm$ 1.1 & 17.8$\pm$ 5.5 & 42.9$\pm$ 34.0 & 10.9$\pm$ 0.2 & 7.3$\pm$ 2.9 & --\\
5302.6 & [Fe XIV] & -- & -- & -- & -- & 6.5$\pm$ 2.9 & -- & -- & --\\
5411.4 & He I & -- & 1.4$\pm$ 0.7 & -- & -- & -- & -- & 2.4$\pm$ 0.3 & --\\
5577.3 & [O I] & 2.9$\pm$ 0.7 & 3.3$\pm$ 0.4 & -- & -- & -- & 1.9$\pm$ 0.5 & 0.2$\pm$ 0.1 & -- \\
5720.7 & [Fe VII] & 11.1$\pm$ 2.0 & -- & -- & 1.4$\pm$ 0.2 & 5.3$\pm$ 2.1 & 2.5$\pm$ 0.7 & 6.9$\pm$ 2.7 & --\\
5754.9 & [N II] & 13.3$\pm$ 2.7 & 4.7$\pm$ 5.8 & 6.4$\pm$ 3.4 & 6.7$\pm$ 2.1 & 6.0$\pm$ 9.6 & 4.9$\pm$ 0.3 & 10.8$\pm$ 0.6 & -- \\
5875.6 & He I & -- & 7.9$\pm$ 0.5 & 8.6$\pm$ 1.1 & 4.2$\pm$ 0.4 & 3.3$\pm$ 1.5 & 3.1$\pm$ 6.1 & 11.4$\pm$ 1.4 & 5.6$\pm$ 2.6\\
6087 & [Fe VII] & -- & 3.6$\pm$ 1.1 & -- & -- & -- & -- & 10.9$\pm$ 0.4 & -- \\
6300.3 & [O I] & 37.6$\pm$ 1.5 & 41.6$\pm$ 8.0 & 17.8$\pm$ 2.7 & 24.6$\pm$ 1.7 & 102.0$\pm$ 2.7 & 7.2$\pm$ 6.4 & 89.6$\pm$ 5.9 & --\\
6363.8 & [O I] & -- & 13.6$\pm$ 1.1 & 4.3$\pm$ 2.5 & 11.8$\pm$ 0.6 & 39.2$\pm$ 5.6 & 2.9$\pm$ 0.5 & 24.7$\pm$ 2.8 & --\\
6374.5 & [Fe X] & -- & 2.5$\pm$ 1.1 & -- & 6.4$\pm$ 0.6 & -- & 1.4$\pm$ 0.2 & 5.1$\pm$ 3.8 & -- \\
6547.9 & [N II] & 115.3$\pm$ 4.7 & 71.8$\pm$ 2.6 & 71.6$\pm$ 3.6 & 69.4$\pm$ 0.9 & 187.6$\pm$ 5.6 & 80.2$\pm$ 2.3 & 46.9$\pm$ 3.5 & 24.9$\pm$ 4.6\\
6562.8 & H$\alpha$ & 286.0$\pm$ 13.4 & 286.0$\pm$ 8.9 & 286.0$\pm$ 8.9 & 286.0$\pm$ 3.4 & 286.0$\pm$ 7.7 & 286.0$\pm$ 10.5 & 286.0$\pm$ 10.5 & 286.0$\pm$ 14.8\\
6583.2 & [N II] & 345.8$\pm$ 14.0 & 215.3$\pm$ 7.7 & 214.7$\pm$ 10.7 & 208.2$\pm$ 2.8 & 562.8$\pm$ 16.7 & 240.6$\pm$ 6.8 & 140.8$\pm$ 10.4 & 74.6$\pm$ 13.8\\
6678.2 & He I & -- & 2.3$\pm$ 0.7 & 4.0$\pm$ 1.2 & -- & -- & 1.2$\pm$ 0.1 & 3.3$\pm$ 1.5 & --\\
6716.4 & [S II] & 69.7$\pm$ 11.8 & 70.2$\pm$ 2.9 & 50.7$\pm$ 1.6 & 45.0$\pm$ 1.7 & 217.4$\pm$ 11.4 & 36.0$\pm$ 0.8 & 32.9$\pm$ 5.2 & 15.0$\pm$ 3.0\\
6730.8 & [S II] & 48.4$\pm$ 16.7 & 68.3$\pm$ 2.6 & 40.7$\pm$ 5.4 & 42.3$\pm$ 1.3 & 158.0$\pm$ 11.4 & 37.8$\pm$ 1.1 & 51.3$\pm$ 5.1 & 16.7$\pm$ 3.8\\
\hline\\
Lambda  & Line ID & MCG-05-23-004 & MCG-05-23-008 & NGC 2992 & MARK 1239 & NGC 3100 & IC 2560 & MCG -06-23-038 & IRAS 11215\\
\hline\\
3586.3 & [Fe VII] & -- & -- & -- & 11.6$\pm$0.7 & -- & 7.5$\pm$1.0 & -- & -- \\
3726,9  & [O II] & 197.3$\pm$3.9 & 117.9$\pm$3.8 & 338.9$\pm$11.2 & 36.8$\pm$5.6 & 155.9$\pm$13.2 & 165.6$\pm$2.9 & 292.1$\pm$14.9 & 296.6$\pm$9.5 \\
3868.8 & [Ne III] & -- & -- & -- & 36.0$\pm$3.3 & -- & 100.2$\pm$7.4 & 58.0$\pm$6.5 & 68.1$\pm$3.8  \\
3889 & H,He I & -- & 39.2$\pm$2.9 & 17.8$\pm$7.1 & 19.7$\pm$3.3 & 24.3$\pm$3.5 & 24.6$\pm$5.9 & 34.1$\pm$20.4 & -- \\
3967,70 & [Ne III],H & 29.1$\pm$2.6 & 25.7$\pm$3.9 & 26.0$\pm$2.4 & 29.6$\pm$1.2 & 19.8$\pm$12.4 & 52.1$\pm$1.7 & 33.4$\pm$4.2 & 34.6$\pm$2.8  \\
4026.2 & He I & -- & -- & -- & 1.8$\pm$0.2 & -- & 2.9$\pm$70.9 & -- & --  \\
4068,76 & [S II] & 33.4$\pm$2.5 & 30.7$\pm$3.2 & -- & 8.6$\pm$1.2 & 14.6$\pm$2.6 & 29.1$\pm$3.0 & 5.1$\pm$0.9 & --  \\
4101.7 & H$\delta$ & -- & -- & 33.7$\pm$2.3 & 29.0$\pm$0.7 & -- & 30.7$\pm$2.7 & -- & 33.8$\pm$3.9  \\
4340.5 & H$\gamma$ & -- & -- & 45.8$\pm$2.9 & 47.3$\pm$1.1 & -- & 51.3$\pm$3.3 & 62.0$\pm$2.4 & 57.3$\pm$4.2  \\
4363.2 & [O III] & -- & -- & 21.5$\pm$2.7 & 17.2$\pm$3.3 & -- & 16.7$\pm$1.7 & 18.2$\pm$19.1 & -- \\
4471.5 & He I & -- & 2.7$\pm$11.8 & -- & 1.3$\pm$0.9 & 8.9$\pm$3.8 & 5.4$\pm$0.3 & 10.3$\pm$3.3 & --  \\
4685.7 & He II & -- & -- & 7.6$\pm$2.7 & 11.0$\pm$10.7 & -- & 26.6$\pm$6.5 & 6.8$\pm$0.6 & 14.3$\pm$1.3  \\
4861.3 & H$\beta$ & 100.0$\pm$3.6 & 100.0$\pm$4.0 & 100.0$\pm$3.3 & 100.0$\pm$1.8 & 100.0$\pm$4.7 & 100.0$\pm$2.0 & 100.0$\pm$3.2 & 100.0$\pm$8.9  \\
4958.9 & [O III] & 41.6$\pm$2.9 & 27.1$\pm$2.5 & 175.3$\pm$3.2 & 57.5$\pm$3.7 & 29.9$\pm$1.5 & 356.4$\pm$3.5 & 238.3$\pm$9.1 & 262.5$\pm$2.5  \\
5006.8 & [O III] & 124.7$\pm$8.8 & 81.2$\pm$7.6 & 525.8$\pm$9.6 & 172.5$\pm$11.0 & 89.7$\pm$4.6 & 1069.1$\pm$10.5 & 714.9$\pm$27.3 & 787.6$\pm$7.4  \\
5198,200 & [N I] & 20.0$\pm$7.5 & 14.4$\pm$7.3 & 9.7$\pm$0.4 & 3.5$\pm$1.4 & 13.6$\pm$2.2 & 15.1$\pm$3.0 & 11.1$\pm$0.8 & -- \\
5302.6 & [Fe XIV] & 3.9$\pm$3.3 & -- & 1.0$\pm$0.6 & 4.9$\pm$0.5 & 3.5$\pm$1.9 & -- & 1.2$\pm$1.2 & 4.7$\pm$4.7  \\
5411.4 & He I & -- & -- & -- & -- -- & 2.8$\pm$0.5 & -- & --  \\
5577.3 & [O I] & 5.7$\pm$2.5 & -- & 2.3$\pm$2.6 & 0.3$\pm$0.1 & 11.0$\pm$2.0 & 0.3$\pm$0.1 & -- & 1.5$\pm$0.4  \\
5720.7 & [Fe VII] & -- & -- & -- & 6.5$\pm$7.4 & -- & 9.4$\pm$1.1 & 5.4$\pm$8.0 & 1.6$\pm$1.6  \\
5754.9 & [N II] & -- & -- & -- & 0.5$\pm$7.4 & 4.9$\pm$3.2 & 6.0$\pm$0.8 & 11.9$\pm$6.0 & 13.5$\pm$5.1  \\
5875.6 & He I & -- & -- & 8.3$\pm$1.4 & 10.0$\pm$0.4 & -- & 10.7$\pm$0.5 & 8.9$\pm$30.9 & 13.0$\pm$3.9  \\
6087 & [Fe VII] & -- & -- & 2.4$\pm$4.2 & 10.3$\pm$1.9 & -- & 15.9$\pm$1.4 & 6.5$\pm$2.9 & --  \\
6300.3 & [O I] & 70.8$\pm$6.4 & 65.4$\pm$4.6 & 27.5$\pm$1.2 & 3.4$\pm$2.2 & 86.3$\pm$4.7 & 45.4$\pm$2.0 & 58.3$\pm$15.5 & 30.5$\pm$ 8.8  \\
6363.8 & [O I] & 28.0$\pm$6.3 & 24.5$\pm$9.3 & 6.3$\pm$0.6 & 5.4$\pm$0.7 & 16.7$\pm$3.9 & 14.9$\pm$0.8 & 19.9$\pm$2.0 & 9.5$\pm$3.3  \\
6374.5 & [Fe X] & 24.3$\pm$3.1 & 25.0$\pm$7.7 & 8.9$\pm$0.5 & 2.1$\pm$0.7 & --& 6.0$\pm$0.7 & -- & 9.6$\pm$4.0  \\
6547.9 & [N II] & 73.5$\pm$6.0 & 120.4$\pm$5.3 & 90.2$\pm$1.1 & 16.2$\pm$1.5 & 89.3$\pm$2.8 & 96.0$\pm$2.5 & 86.9$\pm$4.5 & 53.9$\pm$1.8  \\
6562.8 & H$\alpha$ & 286.0$\pm$19.6 & 286.0$\pm$14.6 & 286.0$\pm$2.9 & 286.0$\pm$5.0 & 286.0$\pm$8.0 & 286.0$\pm$8.2 & 286.0$\pm$14.1 & 286.0$\pm$5.1  \\
6583.2 & [N II] & 220.5$\pm$18.0 & 361.3$\pm$15.9 & 270.5$\pm$3.2 & 48.5$\pm$4.4 & 267.8$\pm$8.5 & 288.1$\pm$7.5 & 260.8$\pm$13.4 & 161.7$\pm$5.5  \\
6678.2 & He I & -- & -- & 0.5$\pm$4.2 & 1.7$\pm$2.1 & -- & 1.6$\pm$1.3 & 2.1$\pm$2.9 & 2.6$\pm$1.9  \\
6716.4 & [S II] & 164.8$\pm$33.3 & 196.9$\pm$30.6 & 52.0$\pm$1.5 & 6.4$\pm$2.0 & 133.6$\pm$18.4 & 74.8$\pm$2.0 & 86.1$\pm$3.7 & 65.5$\pm$3.0  \\
6730.8 & [S II] & 127.2$\pm$32.3 & 168.6$\pm$43.0 & 47.4$\pm$1.1 & 5.4$\pm$2.0 & 77.7$\pm$18.4 & 79.9$\pm$1.6 & 107.7$\pm$3.6 & 68.8$\pm$3.9  \\
\hline\\
Lambda & Line ID & NGC 3783 & NGC 4303 & NGC 4304 & 3C 278 & NGC 5506 & NGC 5597 & NGC 5664 & NGC 5728 \\
\hline\\
3586.3 & [Fe VII] & 23.8$\pm$1.3 & 5.9$\pm$3.9 &  --  &  --  & 0.2$\pm$0.1 & -- & -- & -- \\
3726,9  & [O II] & 72.8$\pm$6.6 & 157.9$\pm$4.5 & 76.8$\pm$5.4 & -- & 264.5$\pm$3.8 & 122.0$\pm$2.2 & 183.9$\pm$4.6 & 232.5$\pm$3.8\\
3868.8 & [Ne III] & 99.9$\pm$3.5 &  --  &  --  &  --  & 58.1$\pm$1.5 & --  & 29.7$\pm$1.4 & 82.6$\pm$10.0\\
3889 & H,He I & 54.5$\pm$14.0 & 45.0$\pm$14.1 & 36.6$\pm$42.6 & --  & 15.6$\pm$19.5 & 17.2$\pm$0.6 & 20.9$\pm$1.2 & 22.6$\pm$5.2\\
3967,70 & [Ne III],H & 36.9$\pm$2.3 & 39.3$\pm$1.2 & 23.9$\pm$5.3 & -- & 31.4$\pm$0.4 & 14.6$\pm$1.1 & 20.8$\pm$3.2 & 40.9$\pm$4.5\\
4026.2 & He I &  --  &  --  &  --  &  --  & 1.5$\pm$0.2 & 1.6$\pm$0.1 &  --  &  -- \\
4068,76 & [S II] & 34.1$\pm$2.7 & 11.2$\pm$2.5 & 22.7$\pm$4.7 & -- & 15.3$\pm$0.7 & 2.4$\pm$0.1 & 7.6$\pm$2.1 & 12.7$\pm$3.5\\
4101.7 & H$\delta$ & 45.9$\pm$4.5 & -- & -- & -- & 25.6$\pm$0.6 & 27.1$\pm$1.5 & 27.0$\pm$4.1 & 25.4$\pm$7.3\\
4340.5 & H$\gamma$ & 49.0$\pm$4.7 & -- & -- & -- & 47.9$\pm$0.9 & 43.9$\pm$1.6 & 51.1$\pm$3.9 & 46.7$\pm$1.6\\
4363.2 & [O III] & 29.0$\pm$2.3 &  --  &  --  &  --  & 11.5$\pm$0.4 & -- & 2.1$\pm$0.7 & 11.4$\pm$1.6\\
4471.5 & He I & 5.2$\pm$1.0 &  --  & 17.5$\pm$1.9 & 20.4$\pm$5.8 & 4.1$\pm$0.5 & 3.7$\pm$0.3 & 3.2$\pm$1.7 & 2.8$\pm$1.7\\
4685.7 & He II & 31.4$\pm$1.0 &  --  &  --  &  --  & 16.4$\pm$0.5 & 3.1$\pm$0.4 & 12.5$\pm$1.1 & 17.1$\pm$0.9\\
4861.3 & H$\beta$ & 100.0$\pm$5.9 & 100.0$\pm$10.0 & 100.0$\pm$2.6 & 100.0$\pm$5.6 & 100.0$\pm$1.0 & 100.0$\pm$2.5 & 100.0$\pm$4.0 & 100.0$\pm$4.6\\
4958.9 & [O III] & 316.1$\pm$2.2 & 30.8$\pm$4.5 & 23.8$\pm$11.3 &  --  & 232.0$\pm$1.6 & 10.7$\pm$0.2 & 98.6$\pm$2.4 & 303.5$\pm$3.9\\
5006.8 & [O III] & 948.2$\pm$6.7 & 92.3$\pm$13.4 & 71.3$\pm$33.9 &  --  & 695.9$\pm$4.8 & 32.1$\pm$0.6 & 295.8$\pm$7.2 & 910.4$\pm$11.8\\
5198,200 & [N I] & 22.5$\pm$4.3 & 12.7$\pm$2.0 & 16.7$\pm$8.8 &  --  & 10.3$\pm$0.4 & 1.6$\pm$0.1 & 8.2$\pm$0.9 & 16.1$\pm$7.5\\
5302.6 & [Fe XIV] & 7.6$\pm$1.2 &  --  & 15.0$\pm$41.3 & -- & 1.7$\pm$2.3 &  --  &  --  &  2.3$\pm$1.2 \\
5411.4 & He I &  --  &  --  &  --  &  --  & 1.3$\pm$0.7 & 0.2$\pm$0.1 &  --  &  -- \\
5577.3 & [O I] & 2.8$\pm$1.5 & 7.5$\pm$1.8 &  --  &  --  & 0.5$\pm$0.1 &  --  &  --  &  0.7$\pm$0.2 \\
5720.7 & [Fe VII] & 25.4$\pm$1.2 & 0.9$\pm$0.7 & -- & --& 0.9$\pm$0.5 &  --  & 4.0$\pm$1.1 & 2.8$\pm$1.0\\
5754.9 & [N II] & 4.8$\pm$82.6 & 0.7$\pm$1.2 & -- & -- & 2.5$\pm$5.8 & 0.5$\pm$0.1 & 5.0$\pm$0.8 & 5.8$\pm$1.4\\
5875.6 & He I & 25.6$\pm$1.9 & 4.3$\pm$41.6 &  --  &  --  & 11.1$\pm$18.0 & 9.1$\pm$0.3 & 9.8$\pm$0.6 & 8.6$\pm$0.5\\
6087 & [Fe VII] & 43.3$\pm$2.0 &  --  &  --  &  --  & 3.4$\pm$0.6 &  --  & 3.6$\pm$0.7 & 5.3$\pm$2.2\\
6300.3 & [O I] & 29.9$\pm$4.8 & 14.4$\pm$1.1 & -- & -- & 45.0$\pm$0.7 & 2.6$\pm$0.1 & 13.9$\pm$1.1 & 38.9$\pm$2.8\\
6363.8 & [O I] & 24.2$\pm$1.1 & 2.5$\pm$14.1 &  --  & -- & 15.2$\pm$0.2 & 1.4$\pm$0.1 & 4.7$\pm$12.3 & 12.2$\pm$0.7\\
6374.5 & [Fe X] & 19.6$\pm$0.8 & -- & -- &  --  & 0.1$\pm$0.5 & 0.7$\pm$0.2 & 3.1$\pm$12.4 & 4.5$\pm$0.7\\
6547.9 & [N II] & 42.8$\pm$2.0 & 62.6$\pm$2.1 & 99.2$\pm$33.4 & 16.1$\pm$5.7 & 88.4$\pm$0.7 & 41.9$\pm$2.6 & 56.9$\pm$3.7 & 117.9$\pm$4.1\\
6562.8 & H$\alpha$ & 286.0$\pm$6.0 & 286.0$\pm$8.5 & 286.0$\pm$33.4 & 286.0$\pm$23.1 & 286.0$\pm$2.2 & 286.0$\pm$5.4 & 286.0$\pm$12.1 & 286.0$\pm$13.4\\
6583.2 & [N II] & 128.5$\pm$6.0 & 187.9$\pm$6.4 & 297.6$\pm$100.3 & 48.3$\pm$17.1 & 265.2$\pm$2.2 & 125.7$\pm$7.8 & 170.7$\pm$11.2 & 353.8$\pm$12.3\\
6678.2 & He I &  --  &  --  &  --  &  --  & 2.8$\pm$1.6 & 2.7$\pm$0.4 & 2.8$\pm$1.3 & 1.9$\pm$1.3\\
6716.4 & [S II] & 32.6$\pm$4.2 & 47.8$\pm$2.8 & 198.1$\pm$45.9 & 51.2$\pm$45.5 & 94.1$\pm$1.6 & 24.8$\pm$0.5 & 58.1$\pm$3.1 & 97.1$\pm$6.0\\
6730.8 & [S II] & 35.7$\pm$2.7 & 36.1$\pm$3.3 & 112.1$\pm$59.7 &  --  & 106.3$\pm$1.2 & 24.1$\pm$0.3 & 43.5$\pm$7.1 & 66.5$\pm$3.5\\
\hline\\
Lambda & Line ID & NGC 5757 & NGC 6000 & ESO 127-G34 & ESO 138-G01 & NGC 6221 & NGC 6300 & FAIRALL 49 & ESO 103-G35 \\
\hline\\
3586.3 & [Fe VII] &  --  & 0.6$\pm$1.1 & 0.2$\pm$0.1 & -- & 1.2$\pm$0.9 & 24.2$\pm$7.3 &  --  &  -- \\
3726,9  & [O II] & 52.1$\pm$0.9 & 76.7$\pm$1.3 & 310.9$\pm$0.6 & 221.2$\pm$2.3 & 102.3$\pm$3.4 & 541.7$\pm$35.7 & 190.3$\pm$2.8 & 319.2$\pm$7.7\\
3868.8 & [Ne III] & -- & 1.0$\pm$0.6 & 86.8$\pm$0.8 & 92.5$\pm$4.6 & 7.5$\pm$2.8 & 104.7$\pm$11.8 & 59.1$\pm$1.0 & 72.8$\pm$23.4\\
3889 & H,He I & 23.2$\pm$1.3 & 15.0$\pm$0.7 & 21.4$\pm$0.6 & 20.4$\pm$3.4 & 19.2$\pm$23.4 & 45.3$\pm$12.5 & 15.6$\pm$0.8 & 26.3$\pm$12.1\\
3967,70 & [Ne III] ,H & -- & 10.3$\pm$6.3 & 42.1$\pm$0.4 & 40.1$\pm$2.4 & 11.5$\pm$1.3 & 127.9$\pm$16.2 & 25.9$\pm$1.5 & 34.6$\pm$12.4\\
4026.2 & He I &  --  &  --  &  --  & 2.2$\pm$0.3 & 1.5$\pm$0.3 &  --  &  --  &  -- \\
4068,76 & [S II] &  --  & 4.9$\pm$1.3 & 16.6$\pm$1.7 & 16.8$\pm$2.5 & 3.0$\pm$0.6 & -- & 28.7$\pm$0.6 &  -- \\
4101.7 & H$\delta$ & 30.4$\pm$3.5 & 28.2$\pm$2.7 & 29.6$\pm$0.6 & 27.0$\pm$1.4 & 30.7$\pm$2.9 & 36.6$\pm$7.2 & 25.3$\pm$1.1 & 4 -- \\
4340.5 & H$\gamma$ & 50.7$\pm$1.6 & 46.9$\pm$3.3 & 50.9$\pm$0.7 & 47.7$\pm$2.5 & 50.5$\pm$12.3 & 49.3$\pm$11.1 & 40.3$\pm$0.8 & 57.7$\pm$4.3\\
4363.2 & [O III] & 0.6$\pm$0.1 &  --  & 11.6$\pm$0.1 & 30.8$\pm$4.1 & 1.0$\pm$0.2 & --& 13.3$\pm$0.5 & 19.6$\pm$13.6\\
4471.5 & He I & 3.0$\pm$0.6 & 2.5$\pm$1.2 & 6.9$\pm$0.2 & 3.9$\pm$0.3 & 3.7$\pm$0.6 & -- & 2.5$\pm$0.2 & 8.6$\pm$13.5\\
4685.7 & He II & 1.3$\pm$0.3 & 0.1$\pm$0.2 & 19.1$\pm$0.2 & 26.7$\pm$1.4 & 2.1$\pm$0.5 & 40.2$\pm$7.4 & 8.3$\pm$2.1 & 8.1$\pm$1.5\\
4861.3 & H$\beta$ & 100.0$\pm$2.1 & 100.0$\pm$1.3 & 100.0$\pm$0.5 & 100.0$\pm$1.6 & 100.0$\pm$3.6 & 100.0$\pm$23.5 & 100.0$\pm$1.9 & 100.0$\pm$2.7\\
4958.9 & [O III] & 4.9$\pm$0.1 & 6.0$\pm$0.3 & 310.2$\pm$1.3 & 274.6$\pm$1.8 & 17.1$\pm$1.8 & 493.5$\pm$3.9 & 126.1$\pm$1.5 & 245.1$\pm$1.5\\
5006.8 & [O III] & 14.7$\pm$0.3 & 17.9$\pm$0.8 & 930.5$\pm$3.9 & 823.9$\pm$5.4 & 51.2$\pm$5.4 & 1480.4$\pm$11.8 & 378.2$\pm$4.5 & 735.4$\pm$4.4\\
5198,200 & [N I] & 3.6$\pm$0.5 & 5.1$\pm$0.6 & 17.0$\pm$4.5 & 4.7$\pm$9.0 & 3.7$\pm$0.2 & 42.0$\pm$8.7 & 11.7$\pm$0.4 & 12.9$\pm$52.5\\
5302.6 & [Fe XIV] & 0.4$\pm$1.2 & 0.3$\pm$0.3 & 2.3$\pm$0.1 & 2.8$\pm$5.7 &  --  & 7.1$\pm$2.3 & 1.1$\pm$1.0 & 2.6$\pm$2.6\\
5411.4 & He I &  --  &  --  &  --  & 2.2$\pm$0.2 & 0.6$\pm$0.5 &  --  &  --  &  -- \\
5577.3 & [O I] & 0.1$\pm$0.1 & 0.1$\pm$0.1 & 0.2$\pm$0.2 & 0.2$\pm$0.2 & 0.5$\pm$0.5 & 1.9$\pm$2.8 &  --  &  -- \\
5720.7 & [Fe VII] &  --  &  --  & 1.0$\pm$1.3 & 9.5$\pm$4.2 & 0.2$\pm$0.1 & 24.1$\pm$5.7 & 3.8$\pm$0.2 &  -- \\
5754.9 & [N II] & 0.6$\pm$0.6 & 1.2$\pm$0.8 & 3.6$\pm$0.2 & 1.7$\pm$5.7 & 0.5$\pm$1.9 & 21.2$\pm$5.2 & 5.6$\pm$0.5 & 13.2$\pm$4.7\\
5875.6 & He I & 6.6$\pm$0.6 & 4.3$\pm$0.8 & 10.9$\pm$0.3 & 8.9$\pm$0.6 & 6.7$\pm$0.6 &  --  & 7.2$\pm$0.3 & 7.8$\pm$1.8\\
6087 & [Fe VII] &  --  &  --  & 5.3$\pm$0.8 & 16.2$\pm$3.4 & 0.5$\pm$0.4 & 5.3$\pm$2.6 & 3.7$\pm$0.4 &  -- \\
6300.3 & [O I] & 4.0$\pm$0.3 & 5.3$\pm$0.4 & 54.4$\pm$1.2 & 30.6$\pm$3.5 & 5.5$\pm$0.3 & 73.2$\pm$5.5 & 27.7$\pm$0.8 & 73.8$\pm$1.2\\
6363.8 & [O I] & 2.9$\pm$0.3 & 3.6$\pm$0.3 & 17.3$\pm$0.3 & 10.2$\pm$1.2 & 2.3$\pm$0.2 & 25.8$\pm$3.1 & 8.8$\pm$0.3 & 23.5$\pm$1.5\\
6374.5 & [Fe X] & 2.2$\pm$0.4 & 2.2$\pm$0.2 & 1.0$\pm$0.2 & 6.9$\pm$1.2 & 1.5$\pm$0.2 & 7.9$\pm$2.7 & 1.7$\pm$0.3 & 2.4$\pm$2.1\\
6547.9 & [N II] & 42.3$\pm$1.8 & 56.1$\pm$1.8 & 124.3$\pm$1.7 & 25.6$\pm$1.8 & 50.1$\pm$3.1 & 165.6$\pm$2.7 & 66.5$\pm$1.2 & 105.8$\pm$0.7\\
6562.8 & H$\alpha$ & 286.0$\pm$4.0 & 286.0$\pm$7.0 & 286.0$\pm$3.5 & 286.0$\pm$5.3 & 286.0$\pm$6.7 & 286.0$\pm$8.1 & 286.0$\pm$2.6 & 286.0$\pm$3.6\\
6583.2 & [N II] & 127.0$\pm$5.3 & 168.3$\pm$5.3 & 372.9$\pm$5.2 & 76.8$\pm$5.3 & 150.4$\pm$9.3 & 496.9$\pm$8.1 & 199.4$\pm$3.5 & 317.4$\pm$2.0\\
6678.2 & He I & 1.4$\pm$0.2 & 1.7$\pm$35.4 & 1.3$\pm$3.0 & 1.8$\pm$1.4 & 1.3$\pm$0.1 &  --  & 4.2$\pm$0.2 & 1.8$\pm$2.7\\
6716.4 & [S II] & 31.6$\pm$0.9 & 26.8$\pm$1.2 & 109.5$\pm$0.8 & 47.4$\pm$6.6 & 30.5$\pm$0.7 & 88.3$\pm$3.7 & 20.5$\pm$0.7 & 79.5$\pm$1.3\\
6730.8 & [S II] & 30.3$\pm$0.6 & 28.9$\pm$1.7 & 126.7$\pm$1.3 & 44.6$\pm$7.6 & 30.6$\pm$1.3 & 82.0$\pm$6.6 & 36.7$\pm$0.6 & 126.9$\pm$1.5\\
\hline\\
Lambda  & Line ID & FAIRALL 51 & NGC 6812 & ESO 339-G11 & NGC 6860 & NGC 6890 & NGC 6915 & NGC 6926 & IC 5063\\
\hline\\
3586.3 & [Fe VII] & 0.4$\pm$0.2 &  --  & 10.0$\pm$1.6 & -- & 14.0$\pm$1.2 &  --  &  --  & -- \\
3726,9 & [O II] & 14.8$\pm$1.1 & 256.7$\pm$5.1 & 239.4$\pm$12.1 & 177.3$\pm$6.7 & 166.4$\pm$4.3 & 235.2$\pm$15.6 & 584.6$\pm$104.7 & 288.8$\pm$6.2\\
3868.8 & [Ne III] & 6.7$\pm$1.9 &  --  & 93.6$\pm$7.5 & 57.1$\pm$7.2 & 134.2$\pm$5.5 &  --  & 18.0$\pm$9.8 & 60.9$\pm$3.3\\
3889 & H,He I & 6.6$\pm$1.9 & -- & 28.1$\pm$6.3 & -- & 38.3$\pm$6.1 & 42.2$\pm$4.7 & 12.5$\pm$5.3 & 22.7$\pm$2.1\\
3967,70 &[Ne III],H & 20.0$\pm$2.9 & 34.1$\pm$7.2 & 24.8$\pm$2.8 & 56.8$\pm$3.3 & 65.4$\pm$1.6 & 24.9$\pm$6.6 &  --  & 34.4$\pm$3.5\\
4026.2 & He I &  --  &  --  &  --  &  --  & 3.5$\pm$1.0 &  --  &  --  & 2.2$\pm$0.3\\
4068,76 & [S II] & 9.6$\pm$1.9 & 20.1$\pm$6.0 & 21.5$\pm$7.6 & -- & -- & -- & -- & 13.1$\pm$1.2\\
4101.7 & H$\delta$ & 25.1$\pm$0.8 & 57.9$\pm$2.8 & 29.7$\pm$2.9 & 44.1$\pm$1.7 & 30.7$\pm$3.3 & -- & 52.5$\pm$12.1 & 30.7$\pm$1.2\\
4340.5 & H$\gamma$ & 34.4$\pm$1.2 & 85.9$\pm$4.6 & 51.8$\pm$1.8 & 59.7$\pm$2.1 & 60.0$\pm$4.4 & 69.4$\pm$6.4 & 79.3$\pm$16.6 & 52.9$\pm$1.9\\
4363.2 & [O III] & 10.1$\pm$1.1 & -- & 14.6$\pm$7.5 & 15.8$\pm$1.2 & 23.4$\pm$3.7 &  --  &  --  & 12.1$\pm$0.8\\
4471.5 & He I &  0.5$\pm$7.2 & 15.9$\pm$1.2 & -- & 3.2$\pm$1.2 & 10.7$\pm$2.1 & 31.6$\pm$4.1 & -- & 6.1$\pm$0.4\\
4685.7 & He II & -- &  --  & 13.0$\pm$1.1 & 6.6$\pm$0.9 & 19.7$\pm$0.7 &  --  & -- & 9.8$\pm$0.4\\
4861.3 & H$\beta$ & 100.0$\pm$2.4 & 100.0$\pm$1.8 & 100.0$\pm$3.3 & 100.0$\pm$5.1 & 100.0$\pm$3.3 & 100.0$\pm$5.3 & 100.0$\pm$18.0 & 100.0$\pm$1.7\\
4958.9 & [O III] & 61.6$\pm$11.8 & 50.7$\pm$0.9 & 262.1$\pm$1.9 & 175.9$\pm$2.7 & 363.6$\pm$1.3 & 37.8$\pm$3.3 & 146.2$\pm$7.1 & 261.1$\pm$2.5\\
5006.8 & [O III] & 184.8$\pm$35.5 & 152.2$\pm$2.6 & 786.2$\pm$5.7 & 527.7$\pm$8.2 & 1090.8$\pm$3.8 & 113.5$\pm$9.8 & 438.7$\pm$21.4 & 783.4$\pm$7.4\\
5198,200 & [N I] & 6.1$\pm$1.1 & 23.2$\pm$12.4 & 23.3$\pm$1.0 & -- & 13.8$\pm$0.5 & -- & -- & 10.5$\pm$0.7\\
5302.6 & [Fe XIV] & -- & -- & 6.2$\pm$2.0 & -- & 4.9$\pm$0.6 & -- & 3.6$\pm$1.4 & --\\
5411.4 & He I &   --  &  --  &  --  &  --  & 5.8$\pm$0.3 &  --  &  --  & 1.7$\pm$1.5\\
5577.3 & [O I] &  2.6$\pm$1.1 &  --  & 4.7$\pm$2.5 &  --  & -- &  --  &  --  & 1.3$\pm$1.1\\
5720.7 & [Fe VII] & -- & --  & -- & -- & -- & -- & -- & 2.3$\pm$0.6\\
5754.9 & [N II] & -- & 30.6$\pm$1.4 & -- & -- & 10.7$\pm$3.7 & -- & -- & --\\
5875.6 & He I &  18.6$\pm$2.3 &  --  & 9.1$\pm$1.2 & 18.3$\pm$1.5 & 6.8$\pm$1.7 & -- & --& 8.6$\pm$0.4\\
6087 & [Fe VII]  & -- &  --  & --  & 5.9$\pm$1.6 & 24.7$\pm$0.8 &  --  &  --  & 3.1$\pm$0.5\\
6300.3 & [O I] & 28.7$\pm$4.9 & 81.6$\pm$2.5 & 49.2$\pm$4.1 & 36.5$\pm$7.6 & 46.0$\pm$2.4 & 79.1$\pm$5.1 & 64.7$\pm$27.3 & 35.8$\pm$5.6\\
6363.8 & [O I] & 10.9$\pm$26.1 & -- & 14.5$\pm$7.8 & 15.9$\pm$1.9 & 16.4$\pm$2.2 & 6.7$\pm$8.4 & 12.7$\pm$8.0 & 12.0$\pm$3.7\\
6374.5 & [Fe X] & -- &  --  & 3.6$\pm$7.7 & 3.9$\pm$2.2 & 12.7$\pm$0.8 & 13.4$\pm$57.6 & 6.3$\pm$6.8 & 2.5$\pm$3.7\\
6547.9 & [N II] & 59.7$\pm$2.7 & 116.2$\pm$2.8 & 132.5$\pm$3.0 & 79.0$\pm$2.0 & 95.1$\pm$3.2 & 173.3$\pm$8.3 & 119.9$\pm$3.1 & 60.6$\pm$2.2\\
6562.8 & H$\alpha$ & 286.0$\pm$8.0 & 286.0$\pm$8.7 & 286.0$\pm$12.5 & 286.0$\pm$6.4 & 286.0$\pm$10.8 & 286.0$\pm$10.1 & 286.0$\pm$14.1 & 286.0$\pm$6.7\\
6583.2 & [N II] & 179.0$\pm$8.2 & 348.6$\pm$8.5 & 397.6$\pm$9.1 & 237.1$\pm$5.9 & 285.3$\pm$9.6 & 519.9$\pm$24.9 & 359.8$\pm$9.4 & 181.9$\pm$6.7\\
6678.2 & He I &  --  & -- & -- &  --  & 0.8$\pm$0.2 &  --  & -- & -- \\
6716.4 & [S II] & 13.8$\pm$1.4 & 192.3$\pm$2.7 & 69.6$\pm$2.4 & 72.1$\pm$7.4 & 43.5$\pm$1.4 & 190.7$\pm$17.5 & 130.8$\pm$9.3 & 77.7$\pm$6.3\\
6730.8 & [S II] & 13.7$\pm$1.4 & 141.9$\pm$7.2 & 64.8$\pm$2.5 & 63.8$\pm$4.6 & 39.8$\pm$1.7 & 146.5$\pm$42.5 & 119.4$\pm$17.0 & 77.5$\pm$2.9\\
\hline \\
Lambda & Line ID & IC1368 &  NGC 7130 &  NGC 7213 &  IC 1459 &  NGC 7469 &  NGC 7496 &  NGC 7591 &  NGC 7582\\
\hline \\
3586.3 & [Fe VII] & -- & -- & -- &  --  & 18.6$\pm$0.9 & -- &  --  & -- \\
3726,9   & [O II] & 456.9$\pm$43.4 & 175.6$\pm$3.0 & 151.7$\pm$9.5 & 124.4$\pm$41.6 & 175.3$\pm$10.9 & 129.5$\pm$0.9 & 336.6$\pm$74.7 & 124.1$\pm$1.7\\
3868.8 & [Ne III] & -- & 68.7$\pm$0.9 & 16.3$\pm$3.9 &  --  & 95.2$\pm$2.9 & 11.1$\pm$0.7 & -- & 32.8$\pm$0.7\\
3889 &  H,He I  &  --  & 24.0$\pm$0.6 & -- & -- & -- & 17.5$\pm$0.5 & -- \\
3967,70  &  [Ne III,H  & -- & 33.7$\pm$0.4 & --& --& 94.0$\pm$5.4 & 15.4$\pm$0.4 & 17.6$\pm$19.7 & 20.5$\pm$2.2\\
4026.2 & He I &  --  & 1.1$\pm$0.5 &  --  &  --  &  --  & 2.6$\pm$0.1 &  --  &  -- \\
4068,76  & [S II] & -- & 17.3$\pm$0.3 & 46.1$\pm$5.7 & 15.2$\pm$2.5 & 20.4$\pm$1.7 & 3.8$\pm$0.2 & -- & 3.3$\pm$0.8\\
4101.7 & H$\delta$ & --& 28.6$\pm$0.9 & -- & -- & -- & 29.7$\pm$0.9 & -- & 25.5$\pm$1.3\\
4340.5 & H$\gamma$ & 56.8$\pm$38.8 & 49.6$\pm$1.6 & 74.7$\pm$3.0 & 118.2$\pm$2.4 & 55.2$\pm$7.4 & 51.8$\pm$1.6 & 58.9$\pm$7.0 & 45.6$\pm$2.5\\
4363.2 & [O III] & -- & 9.6$\pm$0.5 & 19.4$\pm$2.1 & 1.2$\pm$2.0 & 18.9$\pm$2.7 & 2.6$\pm$0.1 &  --  & 2.9$\pm$0.2\\
4471.5 & He I & -- & 6.9$\pm$0.4 & -- & -- & 13.4$\pm$1.1 & 3.6$\pm$0.1 & 7.7$\pm$3.1 & 3.0$\pm$0.1\\
4685.7 & He II & -- & 15.1$\pm$1.1 &  --  &  --  & 26.1$\pm$1.1 & 1.7$\pm$0.1 &  --  & 11.1$\pm$5.0\\
4861.3 & H$\beta$ & 100.0$\pm$5.8 & 100.0$\pm$4.6 & 100.0$\pm$2.6 & 100.0$\pm$31.3 & 100.0$\pm$11.0 & 100.0$\pm$1.0 & 100.0$\pm$7.6 & 100.0$\pm$2.6\\
4958.9 & [O III] & 120.1$\pm$1.6 & 154.2$\pm$1.2 & 40.1$\pm$11.1 & 21.3$\pm$2.0 & 173.7$\pm$2.4 & 18.3$\pm$1.3 & 24.1$\pm$2.2 & 71.6$\pm$3.3\\
5006.8 & [O III] & 360.4$\pm$4.9 & 462.5$\pm$3.6 & 120.2$\pm$33.4 & 63.9$\pm$6.1 & 521.1$\pm$7.3 & 55.0$\pm$3.9 & 72.2$\pm$6.5 & 214.7$\pm$10.0\\
5198,200  & [N I] & 22.7$\pm$12.2 & 16.8$\pm$0.9 & 15.9$\pm$1.8 & 10.8$\pm$29.4 & 22.1$\pm$2.7 & 4.0$\pm$0.4 & 29.6$\pm$5.0 & 5.6$\pm$0.4\\
5302.6 & [Fe XIV] & -- & 3.2$\pm$0.3 & -- & 2.1$\pm$0.9 & 6.2$\pm$0.6 &  --  & 5.0$\pm$5.2 &  -- \\
5411.4 & He I &  --  & 0.8$\pm$0.1 &  --  &  --  &  --  & 0.4$\pm$28.7 &  --  &  -- \\
5577.3 & [O I] & -- &  --  & -- & 1.8$\pm$0.3 &  --  &  --  &  --  & 0.9$\pm$0.1\\
5720.7 & [Fe VII] &  --  & 2.1$\pm$0.1 & -- & -- & 9.5$\pm$0.5 & -- & 12.2$\pm$6.5 & 0.8$\pm$0.2\\
5754.9 & [N II] & -- & 8.3$\pm$1.0 & -- & -- & 8.6$\pm$0.5 & 0.8$\pm$0.2 & 19.8$\pm$2.9 & 1.8$\pm$0.1\\
5875.6 & He I & -- & 5.2$\pm$0.1 & -- &  --  & 19.4$\pm$1.1 & 7.9$\pm$0.2 & -- & 10.7$\pm$1.5\\
6087 & [Fe VII] & -- & 3.3$\pm$0.1 &  --  &  --  & 8.2$\pm$0.7 & -- &  --  & 0.5$\pm$0.3\\
6300.3 & [O I] & 36.3$\pm$4.1 & 29.1$\pm$0.1 & 156.4$\pm$9.8 & 92.9$\pm$9.4 & 16.2$\pm$6.2 & 11.4$\pm$0.2 & 25.0$\pm$4.0 & 8.7$\pm$0.6\\
6363.8 & [O I] & 11.5$\pm$2.6 & 10.2$\pm$0.2 & 47.7$\pm$16.7 & -- & 9.8$\pm$0.6 & 5.0$\pm$0.2 & -- & 3.1$\pm$0.1\\
6374.5 & [Fe X] & -- &  --  & -- & -- & 3.0$\pm$0.6 & -- &  --  & 1.0$\pm$0.1\\
6547.9 & [N II] & 112.9$\pm$3.1 & 121.4$\pm$0.1 & 95.5$\pm$1.6 & 168.7$\pm$6.2 & 57.4$\pm$2.3 & 47.6$\pm$1.4 & 94.8$\pm$4.3 & 62.3$\pm$0.8\\
6562.8 & H$\alpha$ & 286.0$\pm$3.1 & 286.0$\pm$0.2 & 286.0$\pm$5.3 & 286.0$\pm$18.8 & 286.0$\pm$13.0 & 286.0$\pm$3.0 & 286.0$\pm$23.2 & 286.0$\pm$2.2\\
6583.2 & [N II] & 338.6$\pm$9.4 & 364.3$\pm$0.2 & 286.5$\pm$4.7 & 506.0$\pm$18.5 & 172.2$\pm$6.9 & 142.9$\pm$4.3 & 284.4$\pm$12.8 & 186.9$\pm$2.3\\
6678.2 & He I & -- & 3.0$\pm$0.1 &  --  &  --  & 2.5$\pm$0.6 & 2.3$\pm$0.1 & 1.7$\pm$2.9 & 2.4$\pm$0.1\\
6716.4 & [S II] & 70.5$\pm$2.7 & 50.7$\pm$5.1 & 63.1$\pm$15.5 & 116.8$\pm$5.2 & 36.6$\pm$1.4 & 36.7$\pm$0.8 & 75.7$\pm$5.0 & 40.9$\pm$0.9\\
6730.8 & [S II] & 60.5$\pm$2.3 & 50.1$\pm$0.1 & 67.8$\pm$15.5 & 101.3$\pm$5.2 & 35.9$\pm$2.0 & 34.9$\pm$0.4 & 56.9$\pm$4.4 & 38.8$\pm$0.8\\
\hline \\
 Lambda & Line ID & NGC 7590  & NGC 7679 & NGC 7714 &  &  &  &  &  \\
 \hline \\
3586.3 & [Fe VII] & -- & -- &  --  &  &  &  &  &   \\ 
3726,9  & [O II] & 332.3$\pm$19.2 & 119.8$\pm$13.7 & 243.3$\pm$4.7 &  &  &  &  &  \\
3868.8 & [Ne III] & 29.1$\pm$3.2 & 22.2$\pm$2.4 & 9.3$\pm$0.2 &  &  &  &  &   \\
3889 & H,He I & 28.3$\pm$2.6 & 13.5$\pm$3.2 & 21.0$\pm$0.4 &  &  &  &  &   \\
3967,70 & [Ne III],H & 25.7$\pm$2.4 & 12.5$\pm$7.9 & 19.3$\pm$1.1 &  &  &  &  &   \\ 
4026.2 & He I &  --  &  --  & 1.8$\pm$0.1 &  &  &  &  &   \\
4068,76 & [S II] & 13.0$\pm$1.7 & 3.4$\pm$3.3 & 2.8$\pm$0.3 &  &  &  &  &   \\
4101.7 & H$\delta$ & 35.4$\pm$4.2 & 24.4$\pm$1.5 & 29.1$\pm$0.8 &  &  &  &  &   \\
4340.5 & H$\gamma$ & 56.7$\pm$6.0 & 45.6$\pm$2.2 & 50.7$\pm$1.0 &  &  &  &  &   \\
4363.2 & [O III] & 5.2$\pm$2.8 & 11.4$\pm$0.8 & -- &  &  &  &  &   \\ 
4471.5 & He I & 6.7$\pm$1.0 & 2.6$\pm$1.0 & 4.5$\pm$0.1 &  &  &  &  &   \\ 
4685.7 & He II & 7.2$\pm$2.9 & 2.5$\pm$0.2 & 0.8$\pm$0.0 &  &  &  &  &   \\ 
4861.3 & H$\beta$ & 100.0$\pm$6.1 & 100.0$\pm$2.8 & 100.0$\pm$1.3 &  &  &  &  &   \\ 
4958.9 & [O III] & 115.8$\pm$5.5 & 35.1$\pm$4.7 & 46.4$\pm$0.5 &  &  &  &  &   \\ 
5006.8 & [O III] & 347.5$\pm$16.5 & 105.3$\pm$14.0 & 139.2$\pm$1.6 &  &  &  &  &   \\ 
5198,200 & [N I] & 12.7$\pm$1.3 & 8.6$\pm$4.8 & 2.0$\pm$0.0 &  &  &  &  &   \\ 
5302.6 & [Fe XIV] &  --  & -- &  --  &  &  &  &  &   \\ 
5411.4 & He I &  --  &  --  & 0.3$\pm$0.1 &  &  &  &  &   \\
5577.3 & [O I] & 2.9$\pm$0.2 &  --  & -- &  &  &  &  &   \\ 
5720.7 & [Fe VII] &-- & -- &  --  &  &  &  &  &   \\ 
5754.9 & [N II] & 8.0$\pm$1.1 & 4.9$\pm$3.2 & 0.7$\pm$0.1 &  &  &  &  &   \\
5875.6 & He I & 6.7$\pm$2.6 & 12.4$\pm$11.0 & 12.4$\pm$0.2 &  &  &  &  &   \\ 
6087 & [Fe VII] &  --  & -- &  --  &  &  &  &  &   \\
6300.3 & [O I] & 28.5$\pm$3.3 & 8.3$\pm$0.6 & 4.0$\pm$0.2 &  &  &  &  &   \\
6363.8 & [O I] & 7.5$\pm$2.1 & 4.0$\pm$1.4 & 1.7$\pm$0.2 &  &  &  &  &   \\
6374.5 & [Fe X] & 4.1$\pm$1.1 & 2.0$\pm$0.6 & 0.7$\pm$0.2 &  &  &  &  &   \\ 
6547.9 & [N II] & 79.7$\pm$6.3 & 54.6$\pm$4.8 & 32.5$\pm$1.2 &  &  &  &  &   \\
6562.8 & H$\alpha$ & 286.0$\pm$11.9 & 286.0$\pm$13.9 & 286.0$\pm$3.8 &  &  &  &  &   \\ 
6583.2 & [N II] & 239.1$\pm$18.8 & 163.8$\pm$14.3 & 97.6$\pm$3.6 &  &  &  &  &   \\
6678.2 & He I &  --  & 5.9$\pm$2.0 & 4.1$\pm$0.1 &  &  &  &  &   \\ 
6716.4 & [S II] & 98.7$\pm$8.9 & 37.3$\pm$2.2 & 23.4$\pm$1.0 &  &  &  &  &   \\ 
6730.8 & [S II] & 78.4$\pm$4.8 & 28.0$\pm$2.2 & 23.5$\pm$0.7 &  &  &  &  &   \\ 
\enddata
\end{deluxetable}

\end{document}